\documentclass[12pt]{iopart}
\usepackage[utf8]{inputenc}
\usepackage{graphicx}
\usepackage{siunitx} 
\usepackage{wrapfig}
\usepackage{amsmath}
\usepackage[normalem]{ulem}

\begin{document}

\title{Observation and characterisation of trapped electron modes in Wendelstein 7-X}

\author{A. Kr\"amer-Flecken$^1$, J.H.E. Proll$^2$, G. Weir$^2$, P. Costello$^2$,\\ G. Fuchert$^2$, J. Geiger$^2$, S. Heuraux$^3$, A. Knieps$^1$,\\ A. Langenberg$^2$, S. Vaz Mendes$^2$, N. Pablant$^4$, E. Pasch$^2$,\\ K. Rahbarnia$^2$, R. Sabot$^5$, L. Salazar$^3$, H.M. Smith$^2$,\\ H. Thomsen$^2$, T. Windisch$^2$, H.M. Xiang$^1$ and the W7-X-team\footnote{See Klinger et al. 2019 (https://doi.org/10.1088/1741-4326/ab03a7) for the W7-X Team.}}

\address{$^1$Forschungszentrum Jülich GmbH, IFN-1 – Plasma Physics, Jülich, Germany\\
$^2$Max Planck Institut für Plasmaphysik, Greifswald, Germany\\
$^3$Institut Jean Lamour 7198 CNRS, Université de Lorraine, F-54000 Nancy, France\\
$^4$Princeton Plasma Physics Laboratory, Princeton, NJ 08543-0451, USA\\
$^5$CEA, IRFM, F-13108 Saint-Paul-Lez-Durance, France
}   
\date{\today}

\begin{abstract}
    \noindent In the past, quasi coherent modes were reported for nearly all tokamaks. The general definition describes modes as quasi coherent when the magnitude squared coherence is in the range of \SIrange{0.3}{0.6}{}. Quasi coherent modes are observed in the plasma core as well as in the plasma edge and can have quite different physical origins. The one in the core are observed in plasmas with low collisionality, where the electron temperature exceeds the ion temperature in the plasma core. This is the case for electron cyclotron heating in general. The origin of these modes are electrons trapped within a magnetic mirror, as reported in the past from various fusion devices. The so-called trapped-electron modes (TEMs) belong to drift wave instabilities and can be destabilized by electron-temperature gradients in the plasma core. From the diagnostic point of view, quasi coherent modes appear as fluctuations in electron density and temperature. Therefore, the microwave reflectometer is very well suited to monitor these modes.\\This paper describes experiments, conducted at the Wendelstein 7-X stellarator (W7-X), which aim at detecting quasi coherent modes at low wave numbers. A Poloidal Correlation Reflectometer (PCR) installed at W7-X, is able to measure low wave numbers ($k_\perp\le\SI{3.5}{\per\centi\meter}$). For medium line-averaged densities ($\int n_e\le\SI{6e19}{\per\square\meter}$) the plasma core is accessible for this diagnostic. For different magnetic configurations and plasma parameters, broad quasi-coherent  structures are observed in the coherence spectra. From the analysis of the rotation and the poloidal structure, these quasi coherent (QC) modes show the properties of electron-temperature-gradient driven TEMs. A linear relation between the mode velocity and the rotation frequency is found. The relation is uniform and confirms the nature of QC-mode observation as TEM in tokamaks, too. 
\end{abstract}

\section{Introduction}
\noindent Quasi coherent (QC) structures were observed in the power spectra of nearly all tokamaks. The quasi-coherent-term describes the fact that the structures have no monochromatic frequency. Consequently, they cover a large part of the frequency spectrum. The quasi-coherent structures, also called QC-mode, differ in frequency from device to device and covers a frequency range approximately \SIrange{30}{150}{\kilo\hertz}. First measurements of QC-modes are reported from the limiter tokamak T-10~\cite{Vershkov:2004}, measured by correlation reflectometry. At the limiter tokamaks TEXTOR~\cite{akfl:2004} and Tore Supra\cite{arnichand_quasi-coherent_2014} QC-modes have been observed as well by use of poloidal correlation reflectometry (PCR) and Doppler reflectometry (DR), respectively. The modes are observed mostly at the low-field side of tokamaks and tend to disappear at the high-field side~\cite{Vershkov:2010}. Experiments at JET~\cite{arnichand_discriminating_2015} and ASDEX Upgrade~\cite{Vanovac:2023} have reported observations of QC-modes in divertor plasmas as well. Also, at HL2-A and J-TEXT~\cite{Zhong:2016} QC-modes are detected in density fluctuation spectra. Also in the Large Helical Device, evidence for QC-modes in the plasma core is found~\cite{Nakata:2019}. Most recently, KSTAR~\cite{lee_observation_2018, lee_study_2020} reported about the measurement of QC-modes by microwave imaging reflectometry (MIR). Beside the measurement of QC-modes in density fluctuation spectra, measurements with heavy ion beam probes (HIBP) are performed at T-10~\cite{drabinskiy_radial_2019}. It demonstrates that the QC-mode is also visible in potential fluctuation spectra, which are related to the radial electric field ($E_r$) and the perpendicular velocity ($v_\perp$). QC-modes are also observed as temperature fluctuations using CECE as described in \cite{Vanovac:2023}. QC-modes are observed in different radial regions and for a variety of plasma parameters, pointing towards different physical mechanisms generating the QC-modes. This paper concentrates on QC-modes observed in the plasma core and for conditions where the electron temperature $T_e$ is larger than the ion temperature $T_i$. As far as calculated in the existing literature, the obtained wave number of those QC-modes is below $k_\perp\le\SI{3}{\per\centi\meter}$. At Tore Supra, the modes are found in the linear ohmic confinement regime (LOC) and they disappear in the saturated ohmic confinement regime (SOC), which implies that the QC-mode is stabilized with higher plasma density and collisionality, respectively. Nonlinear gyrokinetic simulations and a synthetic diagnostic characterizes the QC-mode as trapped electron mode (TEM)~\cite{arnichand_identification_2016}. TEMs, a type of drift wave instability, can play a significant role in energy and particle transport~\cite{Liewer:1985} in fusion plasmas. In general, TEMs are driven by the normalized gradients of the background plasma, $R/L_{T_e}$ and $R/L_n$~\cite{Nordman:1990, Dannert:2005}, where $R$ denotes the radius and $L$ the scale length of the quantity noted in the index. TEMs connected with the observation of QC-modes in the plasma core are mostly electron-temperature-gradient-driven TEMs and observed best in plasmas where the electron component is predominately heated, as is the case of ECRH heated plasmas. In these plasmas, the ions stay rather cold and ion-temperature-gradient (ITG)-modes are linearly stable, while TEMs are not. The driving mechanism for these TEMs is equivalent to that of toroidal ITG, except that the role of ions and trapped electrons have been exchanged~\cite{Kim:2021}.\\At W7-X~\cite{Bosch:2013}, a quasi-isodynamic stellarator, neoclassical transport has been minimized and the transport of heat and particles is mainly dominated by turbulence. Gyrokinetic simulations predict weak TEM activity on the outboard side~\cite{proll_collisionless_2013} which is mostly driven by ions. However, W7-X is stable against density-gradient-driven TEMs as shown by Proll and co-workers~\cite{Proll:2012, Proll:2013}. Instead, a variant of the universal instability arises at long wavelengths~\cite{Helander:2015, Costello:2023} and an ion-driven TEM at short ones~\cite{podavini:2024}. Recently, a paper by Wilms and coworkers~\cite{wilms:2024} discusses the conditions for the existence of $\nabla T_e$ driven TEMs in W7-X. In this paper, predictions for turbulent electron-heat-flux driven by trapped-electron-mode (TEM) and electron-temperature-gradient (ETG) turbulence in the core of the plasma are reported.\\With this background in mind, experiments have been carried out in different magnetic configurations at W7-X to search for the existence of QC-modes which are driven by $\nabla T_e$ TEMs. In case these QC-modes are observed, it is of interest to find a relation between them and the reported QC-modes in literature. The paper is organized as follows: Section~\ref{Exp} describes the experimental set-up and the diagnostic used. In section~\ref{Res} the results of the measurements for different magnetic configuration are presented, with emphasis on the mode rotation and the structure size. Section~\ref{Rel} combines the results from W7-X and develops a linear relation for the estimation of the mode velocity from the measured frequency of the QC-mode. Furthermore, an extrapolation to reported QC-mode observations at Tore Supra and TEXTOR is performed. It suggests that the scaling is not only valid for W7-X, but, can be applied for other devices, including tokamaks. Section~\ref{Sum} summarizes the results and develops conclusions for further analysis. 

\section{Description of Experiments and Methodology}\label{Exp}
\noindent The investigations outlined below are performed at the superconducting nearly quasi-isodynamic stellarator W7-X. It has a 5-fold symmetry and consists of 10 non-planar coils and 4 planar coils in each of the 5 modules~\cite{Bosch:2013}. With this coil set, different magnetic configurations are realized characterized by the rotational transform ($\iota$) at the LCFS: from configurations with low iota ($\iota_{LCFS}=0.856$) to high iota ($\iota_{LCFS}=1.178$), where LCFS denotes the last closed flux surface. W7-X is equipped with up to \SI{10}{\mega\watt} electron-cyclotron heating and \SI{3.5}{\mega\watt} neutral-beam injection using hydrogen beams. Numerous diagnostics are installed, from which only those are mentioned that are used in this presentation. A set of Mirnov coils~\cite{endler:2015} located in the 2$^{nd}$ half of module 1 is implemented for MHD- and Alfv\'en-mode studies up to \SI{800}{\kilo\hertz}. For the measurement of electron density and temperature profiles, a Thomson Scattering diagnostic is available~\cite{Pasch:2016a}. The turbulence rotation is determined from Doppler Reflectometry (DR)~\cite{Estrada:2021} and Poloidal Correlation Reflectometry (PCR)~\cite{akfl:2017a, Windisch:2017}. The latter is operated in O-mode and covers the Ka- and U-band frequencies. For O-mode operation, the frequency range of both bands corresponds to an electron-density range of \SIrange{0.6e19}{4.5e19}{\per\cubic\meter}. The instrument is capable to resolve wave number in the range \SIrange{0}{3.5}{\per\centi\meter} as demonstrated in \cite{SOLDATOV:2009}. The launching antenna and the 4 receiving antennae have a common focal point and allow for 6 different antenna combinations probing a poloidal distance of \SIrange{7}{35}{\milli\meter} in the region of the plasma, where steep density- and temperature gradients are observed. Due to the common focal point, the poloidal distance decreases towards the plasma core.

\begin{figure}
    \centering
    \includegraphics[scale=0.8]{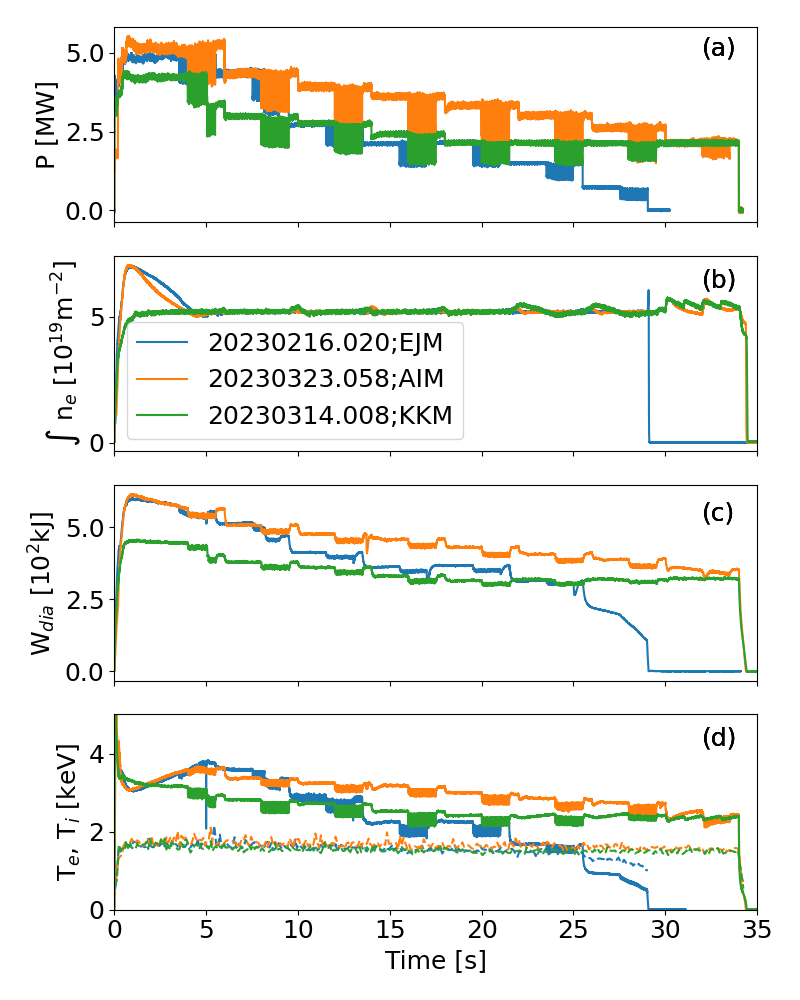}
    \caption{The important parameters of the investigated discharges as function of time. (a) ECRH-power, (b) line averaged density, (c) diamagnetic energy and (d) central electron (solid) and ion (dashed) temperature}
    \label{fig:Overview}
\end{figure}

\begin{figure}
    \centering
    \includegraphics[scale=0.8]{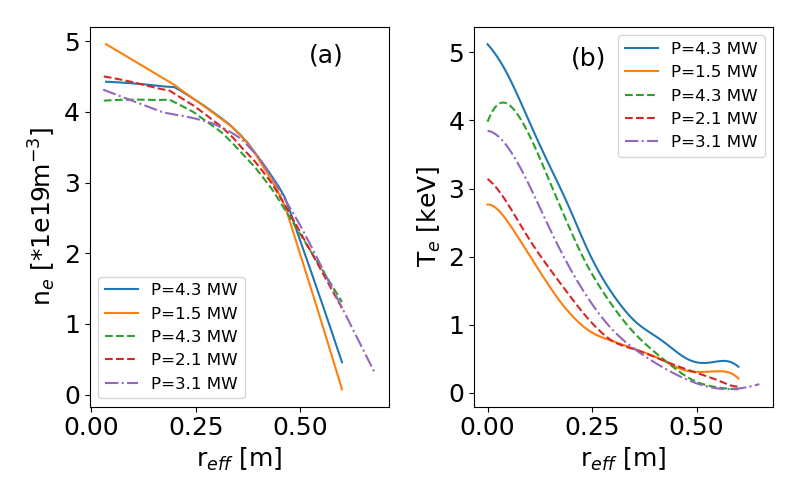}
    \caption{(\textbf{a}) Electron density- and (\textbf{b}) temperature-profiles for the programs in the low mirror configuration \textit{20230216.020} (solid lines) and for the standard configuration \textit{20230323.058} (dashed lines) for the highest and lowest power level. In addition, the profiles for the high mirror configuration \textit{20230314.008} (dash dotted line) at $P_{ECRH}=\SI{3.1}{\mega\watt}$ is added.} 
    \label{fig:profiles}
\end{figure}

\noindent As mentioned in the introduction, the influence of the magnetic configuration on the TEMs and the QC modes caused by them is the object of the investigation. The magnetic configuration can be varied via modifying the current in the non-planar and planar coils. The main experimental programs discussed here are performed in (i) the so-called standard configuration (EIM -- 20230323.058) and (ii) a low-mirror configuration (AIM -- 20230216.020, .036), both having $\iota_{LCFS}=0.97$, according to the vacuum equilibrium calculation~\cite{Andreeva:2002}. The calculated mirror ratio for both discharges is \SI{5.5}{\percent}. Furthermore, a high-mirror configuration (KKM -- 20230314.008) with a mirror ratio of \SI{8.8}{\percent} is investigated. All programs have similar line averaged density of $\int n_e dl$=\SI{5.2e19}{\per\square\meter} as seen in fig.~\ref{fig:Overview}(b). Due to the decreasing power level (fig.~\ref{fig:Overview}(a)), the diamagnetic energy (fig.~\ref{fig:Overview}(c)) and the central electron temperature, measured by the ECE-diagnostic, are decreasing. The central ion temperature ($T_i$), measured by an X-ray Imaging Crystal Spectrometer (XICS) is constant at $T_i\approx\SI{1.8}{keV}$ and independent of $P_{ECRH}$. Each power step has a time interval where the power is modulated for heat-pulse-propagation studies. These results are discussed elsewhere. For the presented analysis, the electron temperature and density profiles are calculated from Thomson Scattering. The density profiles are normalized with the line averaged density from the interferometer~\cite{Knauer:2016}. To reduce the scatter in the data, a locally weighted regression (LOWESS) is applied. While the electron-density ($n_e$) profiles are constant during the discharge, the electron-temperature ($T_e$) profiles change their shape and gradient with decreasing power. For all programs, the profiles are shown in fig.~\ref{fig:profiles}.

\noindent For searching QC-modes in the plasma core, the PCR system is set to a frequency-hopping operation with frequency steps of \SI{0.5}{\giga\hertz} lasting \SI{10}{\milli\second} for each step. Two frequency scans are performed. The first one lasts for \SI{370}{\milli\second}, followed by a 2$^{nd}$ scan of \SI{100}{\milli\second}, used for radial correlation measurements around a fixed density of \SI{2e19}{\per\cubic\meter}. Both scans were repeated one after the other until the program has ended. The in-line and quadrature component of each band is digitized with a sampling frequency of \SI{5}{\mega\hertz}. In the further analysis, the U-band is used, because it probes the plasma core. In a first analysis step, the cross power spectrum (CPSD) is estimated from two complex time series $x(t), y(t)$, representing two antennae from the antenna array. Furthermore, the magnitude squared coherence ($\gamma_{xy}^2$) for this antenna combination is estimated from the CPSD ($S_{xy}(f)$) and the power spectral density of each antenna signal $S_{xx}(f)$ and $S_{yy}(f)$, respectively, as:\begin{equation}\label{Coherence}
     \gamma_{xy}^2(f) = \frac{S_{xy}(f)^2}{S_{xx}(f)\, S_{yy}(f)}
\end{equation} In the rest of the paper, the term coherence is always used for $\gamma_{xy}^2(f)$. The coherence is calculated for windows of \SI{1}{\milli\second} and averaged over 5 windows. Within one frequency step 2 coherence spectra are obtained. Furthermore, a sliding average in frequency across 3 consecutive steps is applied for noise reduction.\\Beside the detection of the QC-mode structure in the coherence spectra, the estimation of the propagation time ($\Delta t$) from cross correlation analysis is applied. The propagation time is defined as\begin{equation}\label{equ:delta-t}
    \Delta t = \underset{t}{\arg\max}\left( (|x \star y|)(t)\right)
\end{equation}where $x, y$ denote the time series from different antenna and $\star$ denotes the cross correlation operator. The turbulence velocity perpendicular to the magnetic field ($v_\perp$) is then calculated by making use of the well-proven elliptical model~\cite{Briggs:1950, He:2016, Han:2021}, which yields
\begin{equation}\label{equ:vperp}
    v_\perp = \frac{\Delta z \, \Delta t}{\Delta t^2 - \tau^2}
\end{equation} Here, $\Delta z$ denotes the perpendicular distance with respect to the magnetic field between the antennae and $\tau$ denotes the time lag, where the maximum of the cross correlation function (CCF) equals the auto correlation function (ACF); $max(CCF(\tau)) = ACF(\tau)$. The perpendicular turbulence velocity is calculated for each combination, and a mean $v_\perp$ is estimated by averaging across all antenna combinations, except the one with the smallest poloidal distance. This combination includes very fast decaying turbulence rotating at a different velocity, compared to the other combinations. The estimation of $v_\perp$ can be frequency selective and allows measuring the velocity for mode structures in a certain frequency interval as determined from the coherence spectra. In general, the measured turbulence velocity is a combination of the $E\times B$-velocity and an additional phase velocity ($v_\perp = v_{E\times B} + v_\Phi$). For low frequency modes e.g. low mode number MHD-modes it is known that they rotate at the same speed as the plasma \cite{Vershkov:1999a} and $v_\Phi$ can be neglected. For high frequency modes $v_\Phi$ may be non-negligible.

\section{Experimental Observations}\label{Res}
\noindent In this section, the identification of QC-modes within the PCR-diagnostic is described. Furthermore, the investigations for the program 20230323.058 regarding the mode velocity (subsection~\ref{QC-Velocity}) and its properties (subsection~\ref{QC-Properties}) are presented. The findings are compared against gyrokinetic calculations for this magnetic configuration. Furthermore, the effect of the magnetic configuration will be presented and discussed (subsection~\ref{MagnConf}). Together, all findings from the observed QC-modes are in agreement with the $\nabla T_e$-driven TEMs.  At the end of this section a scaling of the mode frequency as function of the mode rotation will be presented as well as its extrapolation to tokamaks experiments, where $\nabla T_e$-driven TEMs have been observed, too.

\subsection{Identification of QC-modes}\label{QC-Identification}
\begin{figure}
    \centering
    \includegraphics[scale=0.5]{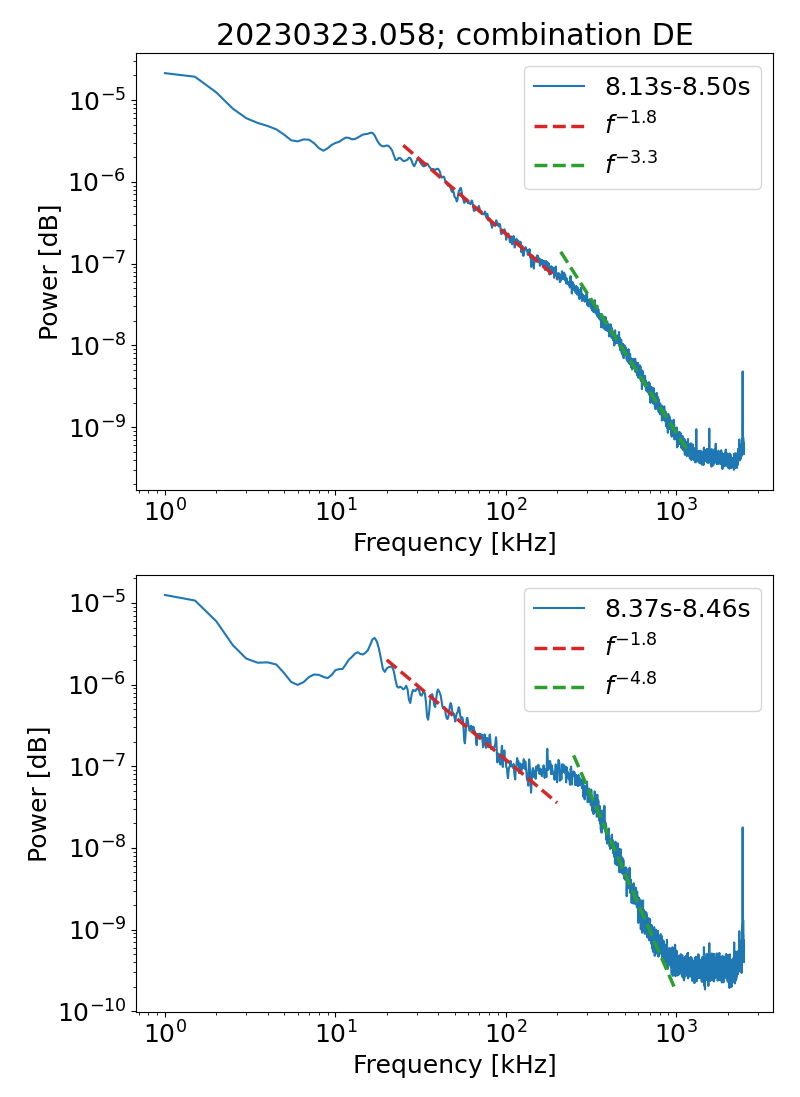}
    \caption{\textbf{(a)}: Cross power spectrum for the antenna combination DE for the full scan. Clearly seen are the two different power laws for the decay of the power spectrum. \textbf{(b)}: Cross power spectrum for the part of the scan which probes the plasma core In between the two power laws describing the decay of the spectrum, a region with injection of energy is visible. This is supposed due to the QC-mode.}
    \label{fig:CPSD}
\end{figure}

For a scan of the PCR lasting for \SIrange{8.13}{8.50}{\second} at a power level of $P_{ECRH}=\SI{4.2}{\mega\watt}$ the CPSD is calculated (see Fig.~\ref{fig:CPSD}a). The figure shows the spectrum obtained for the full time of the scan, representing the radial range from the SOL to the plasma core. The decay of the spectrum for $f\ge\SI{20}{\kilo\hertz}$ is displayed by two power laws. For the range \SIrange{25}{200}{\kilo\hertz} the decay follows $S\propto f^{-1.8}$ and for the range \SIrange{280}{1000}{\kilo\hertz} a decay of $S\propto f^{-3.3}$ is found. The figure~\ref{fig:CPSD}b is observed for a time interval (\SIrange{8.37}{8.47}{\second}) when the core plasma is probed. The spectrum is represented by the same power law for the low frequency range, and for the high frequency range the decay is much steeper $S\propto f^{-4.8}$. However, for the frequency interval, \SIrange{140}{280}{\kilo\hertz} a hump in the spectrum is observed. It indicates an injection of additional energy into the system. The underlying structure is supposed to be the QC-mode. This structure is better visualized when the coherence Eqn.~\ref{Coherence} is calculated. 

\begin{figure}
    \centering
    \includegraphics[scale=0.6]{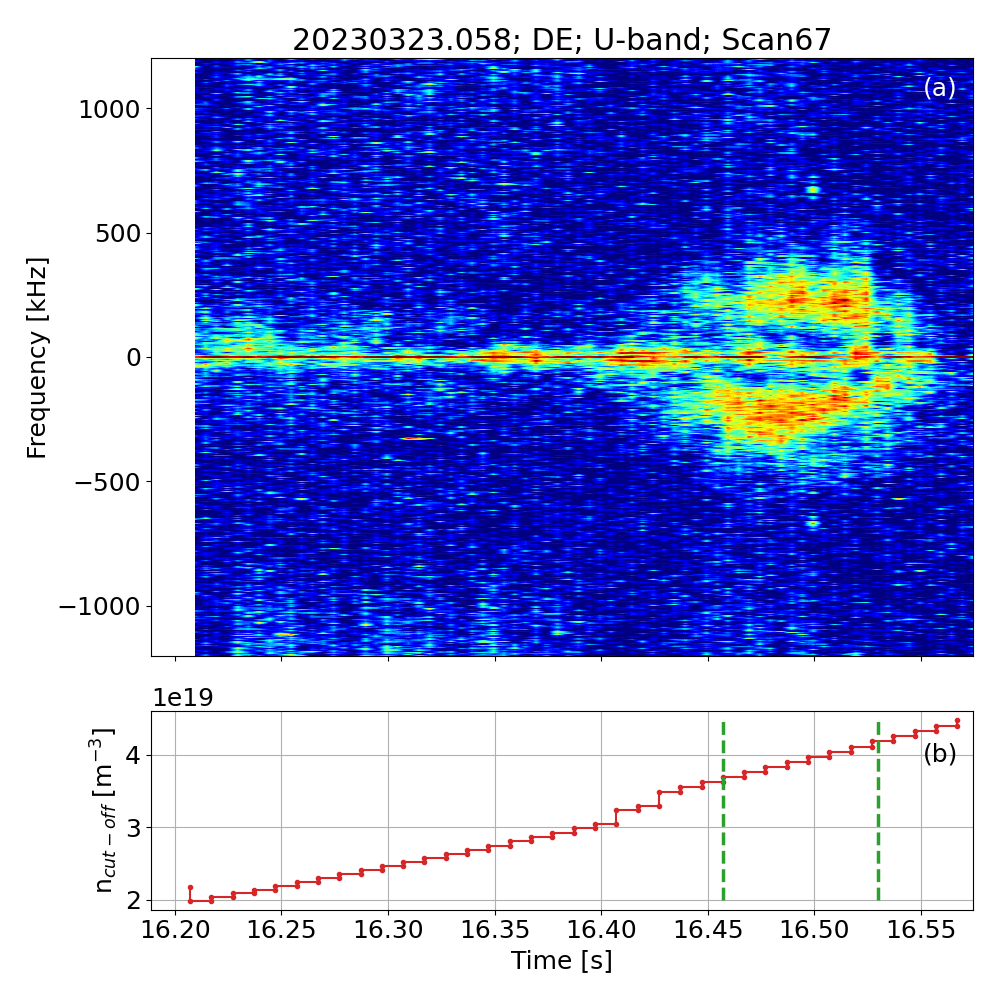}
    \caption{Coherence spectrogram (a) for one scan of the PCR diagnostic, showing the broad quasi QC-mode structure and the related electron density. The coherence spectrum is calculated for combination DE with $\Delta z = \SI{17}{\milli\meter}$. (b) of the reflectometer scan. The green dashed lines denote the density range where the QC-mode is observed.}
    \label{fig:CoherenceSpectrum}
\end{figure}

\noindent An example for the coherence spectrogram for one scan probing a distance of \SI{17}{\milli\meter} is shown in fig.~\ref{fig:CoherenceSpectrum}a. A broad structure is obtained for the time interval \SIrange{16.46}{16.53}{\second}. The centre of this structure is located at $f_c\approx\SI{220}{\kilo\hertz}$ and it has a half width of $\approx\SI{100}{\kilo\hertz}$, which fulfils the requirements for a quasi coherent mode. As can be seen in fig.~\ref{fig:CoherenceSpectrum}b the dashed green lines mark the time interval where the QC-mode is detected in fig.~\ref{fig:CoherenceSpectrum}a. This time interval corresponds to a local density range of \SIrange{3.7e19}{4.2e19}{\per\cubic\meter}. Up to a local density of \SI{3.7e19}{\per\cubic\meter} the coherence spectrogram is empty, besides a low-frequency mode at $\approx$\SI{16}{\kilo\hertz}. This mode rises in the vicinity of the LCFS and is visible up to the plasma core~\cite{han_experimental_2019}. Above a density of \SI{4.4e19}{\per\cubic\meter} the reflection in the plasma is no longer visible, either due to a shallow density gradient and/or because the probing frequency overcomes the maximum density of the discharge at that time.\\In a further step, a decomposition of frequency spectra obtained from turbulent time series, dating back to the work at T-10~\cite{Vershkov:2005}, is used. For W7-X this method is applicable to the coherence spectrum, because the coherence spectrum is free from uncorrelated background and the different components contributing to the spectrum are well distinguishable. Extracting the coherence spectrum for the interval with QC-mode activity (see fig.~\ref{fig:Decomposition}) displays three different structures. Centred at $f=\SI{0}{\kilo\hertz}$ a narrow structure with a half width of \SI{50}{\kilo\hertz} is visible, including the low-frequency mode at $\pm$\SI{16}{\kilo\hertz}. On the positive and negative frequency branch, broad QC-modes centred at $|f_c|\approx\SI{200}{\kilo\hertz}$ are observed with a half width at half maximum (HWHM) of $\approx \SI{200}{\kilo\hertz}$. Each of the three structures is well approximated by a Gaussian shape. The coherence spectrum in the range \SIrange{-750}{750}{\kilo\hertz} can be very well described by the three components, with only three parameters for each component. The different components of the coherence spectrum are denoted by dashed lines and the sum of them is represented by the black dashed line, which describes the measured spectrum well.

\subsection{Estimation of QC-mode rotation}\label{QC-Velocity}
\begin{figure}
    \centering
    \includegraphics[scale=0.8]{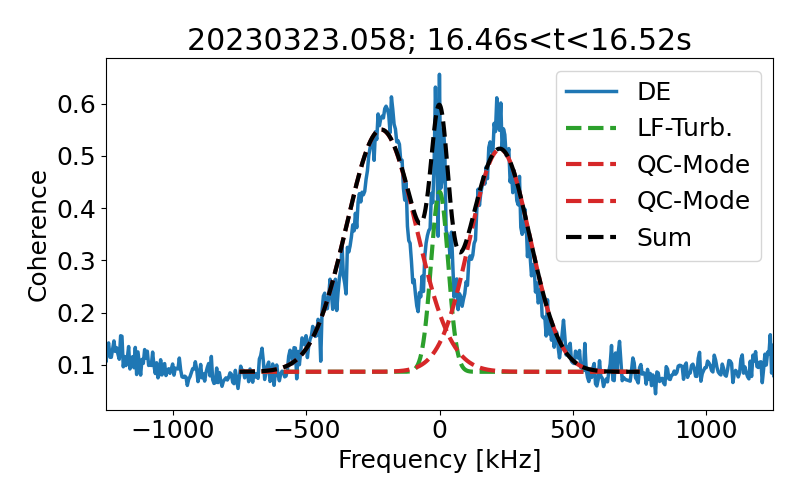}
    \caption{Decomposition of the coherence spectrum: The coherence spectrum is calculated for combination DE with $\Delta z = \SI{17}{\milli\meter}$. The spectrum consists of three Gaussian functions, one describing the low frequency turbulence (LF-Turb) centred around $f\approx$\SI{0}{\kilo\hertz} and two Gaussian functions describing the QC-mode. The sum of the Gaussian components is in good agreement with the measured spectrum}
    \label{fig:Decomposition}
\end{figure}

\begin{figure}
    \centering
    \includegraphics[scale=0.5]{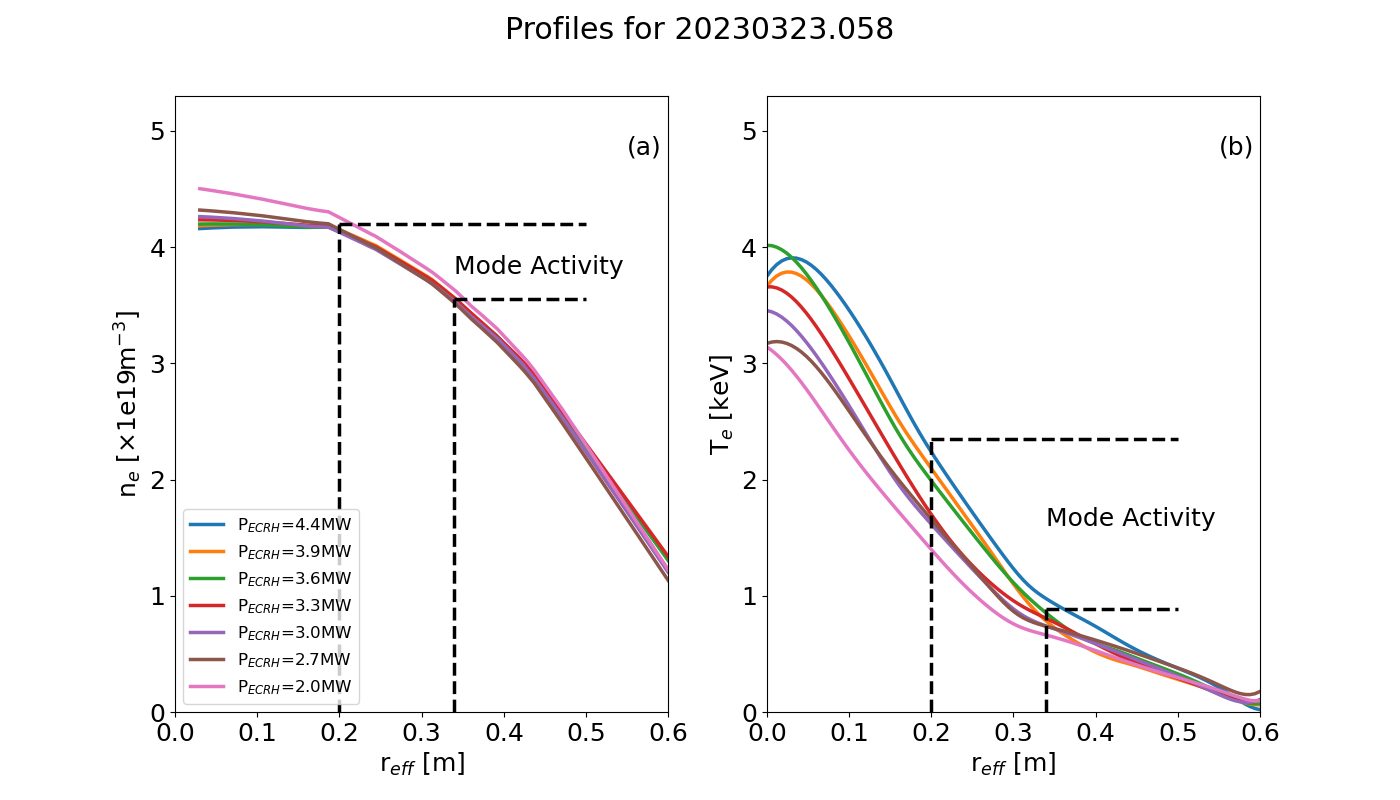}
    \caption{\textbf{a}: Fitted Density profiles from Thomson Scattering for different power levels for program \textit{20230323.058} (solid lines), showing that the shape and the maximum is independent of the power. \textbf{b}: The corresponding electron temperature profiles showing s steepening of the gradient with increasing power level. The dashed horizontal lines indicate the density region where the QC-mode is observed.}
    \label{fig:230323058ne-prof}
\end{figure}

\begin{figure}
    \centering
    \includegraphics[scale=0.65]{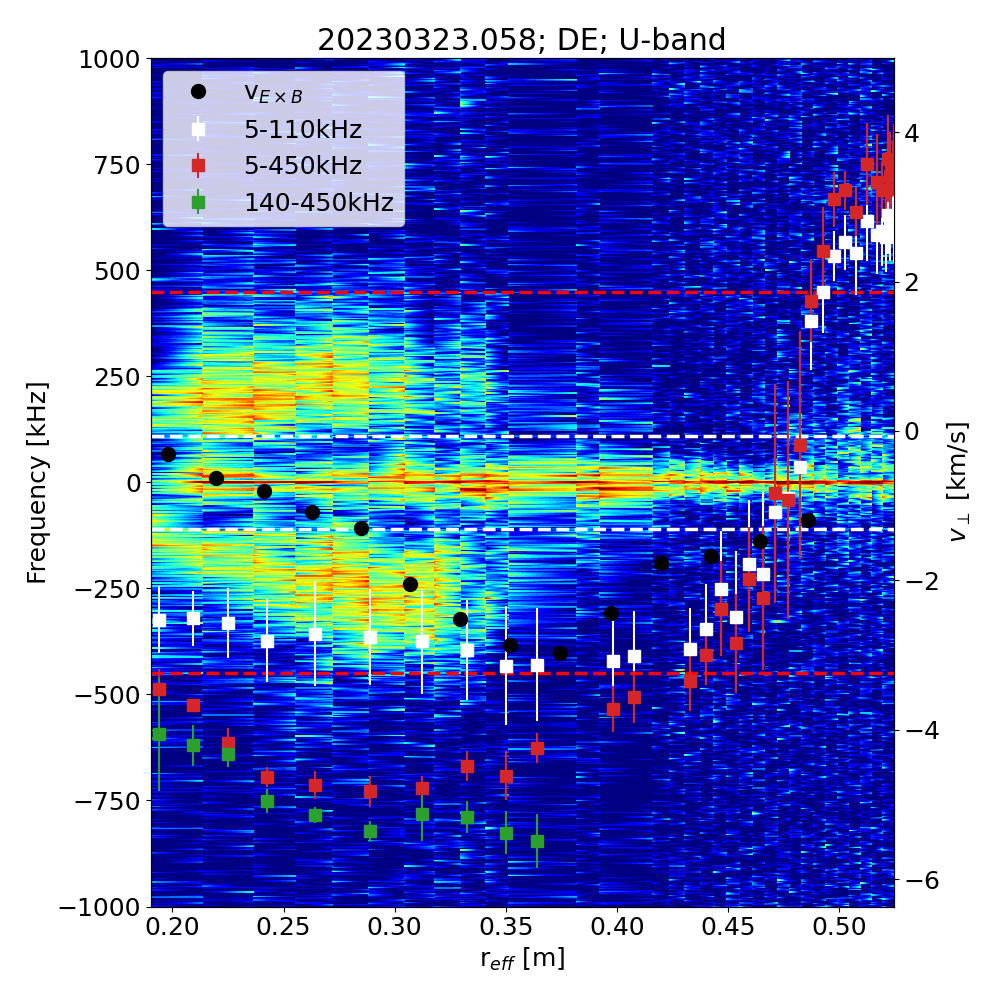}
    \caption{Spectrogram of the coherence for a combination with \SI{15}{\milli\meter} poloidal distance mapped to $r_{eff}$. In addition, the velocities for different frequency intervals are shown. Dashed lines indicate the turbulence contribution, which is taken into account for the velocity estimation. A clear deviation is observed with the onset of the QC-mode activity. The black circles indicate the $v_{E\times b}$-velocity calculated from the mean profiles for this power step.}
    \label{fig:Scan26}
\end{figure}

\begin{figure}
    \centering
    \includegraphics[scale=0.65]{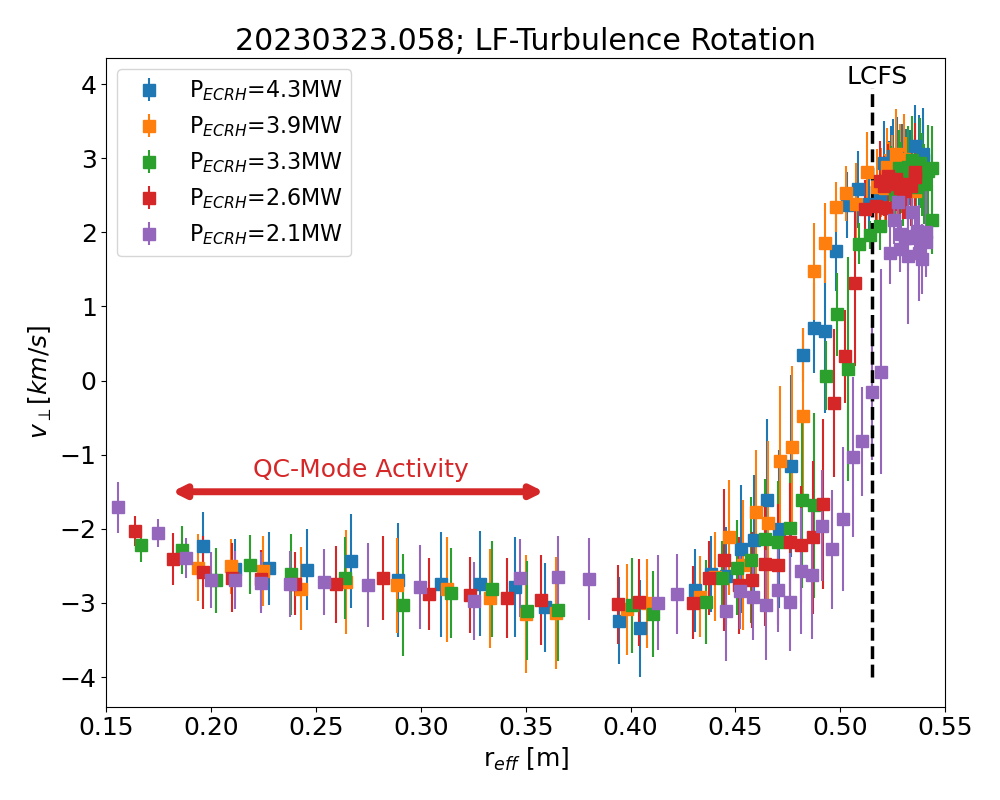}
    \caption{Velocity profiles for the low frequency turbulence interval, \SIrange{5}{110}{\kilo\hertz} including the $E\times B$-rotation. The scan covers a part of the scrape of layer with positive velocities ($r_{eff}\ge$\SI{0.5}{\meter}), the velocity shear region and the plasma core up to $r_{eff}\approx$\SI{0.15}{\meter}. In the radial range with the QC-mode activity, no variation in $v_{E\times B}$ with $P_{ECRH}$ is observed.}
    \label{fig:ExB-rotation}
\end{figure}

\begin{figure}
    \centering
    \includegraphics[scale=0.65]{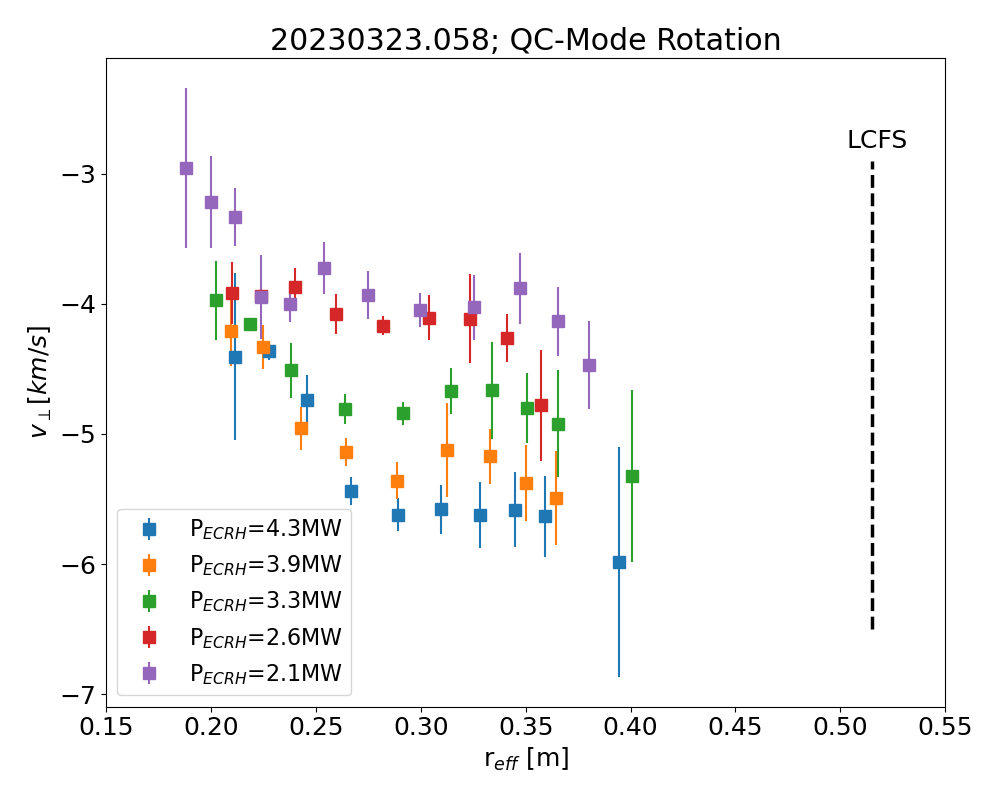}
    \caption{QC-mode rotation for the program \textit{20230323.058} as function of $P_{ECRH}$ and for the selected frequency range. A clear trend with the injected ECRH-power is observed. The measured velocities range from $\SI{-5.6}{\kilo\meter\per\second}\le v_\perp\le \SI{-4}{\kilo\meter\per\second}$ and indicate a rotation in electron diamagnetic drift direction.}
    \label{fig:QC-rotation}
\end{figure}

\begin{figure}
    \centering
    \includegraphics[scale=0.65]{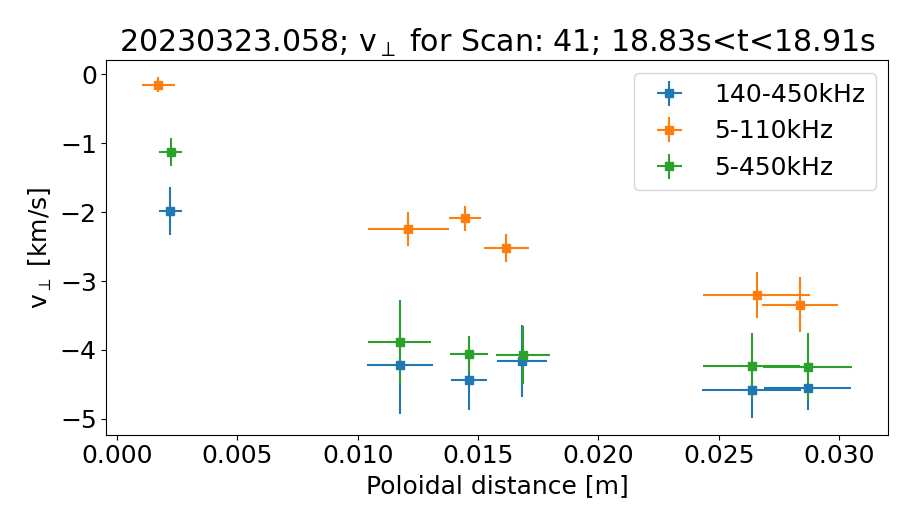}
    \caption{Velocity for all combinations as function of the poloidal distance for the time interval, where QC-mode activity is observed. The velocity is calculated for three different frequency intervals, which include the QC-mode frequency, the low frequency turbulence and the combination of both.}
    \label{fig:vperp-vs-distance}
\end{figure}

\noindent The radial position of the QC-modes is determined from the $n_e$-profiles deduced from Thomson Scattering diagnostic~\cite{Pasch:2016a}. For the standard configuration, the $n_e$-profiles for all power steps are shown in fig~\ref{fig:230323058ne-prof}a. The radial position where the QC-mode dominates the coherence spectrum is within \SI{0.2}{\meter}$\le r_{eff}\le$\SI{0.35}{\meter}. In fig~\ref{fig:230323058ne-prof}b the corresponding temperature profiles are presented. For $r_{eff}\ge\SI{0.35}{\meter}$ the profiles are similar for all power steps and for $r_{eff}\le\SI{0.35}{\meter}$ the electron temperature gradients increases with the power. The estimation of the QC-mode rotation is performed with the cross-correlation analysis, according to Eqn.~\ref{equ:delta-t} and Eqn.~\ref{equ:vperp}, for different antenna combinations. From fig.~\ref{fig:Decomposition} the QC-mode and the low frequency turbulence, including the $E\times B$-rotation, can be discriminated in frequency. The frequency range \SIrange{-110}{110}{\kilo\hertz} includes the $E\times B$-rotation and MHD-activity, where the latter is supposed to rotate at the same speed as the plasma. For the QC-mode rotation, the frequency range \SIrange{140}{450}{\kilo\hertz} and \SIrange{-450}{-140}{\kilo\hertz} is selected, respectively. Both frequency intervals are disjunct and a mixing of components is minimized. The relation between the QC-mode activity and the measured velocities is shown in fig.~\ref{fig:Scan26}. The coherence spectrogram is mapped to $r_{eff}$ using the density profile from TS. The QC-mode activity starts at $r_{eff}\approx\SI{0.36}{\meter}$. Overlaid are the velocity estimations for three different frequency intervals. To indicate which part of the coherence spectrogram contributes to the velocity, coloured dashed horizontal lines are shown, white dashed lines for the frequency interval \SIrange{5}{110}{\kilo\hertz} and red dashed lines for the frequency interval \SIrange{5}{450}{\kilo\hertz}. The velocity estimation for \SIrange{5}{110}{\kilo\hertz} and \SIrange{5}{450}{\kilo\hertz} agree well at $r_{eff}\ge\SI{0.4}{\meter}$. For $r_{eff}\le\SI{0.36}{\meter}$ they start to deviate due to the strong QC-mode activity. With decreasing QC-mode activity, both velocity tend to become equal again. Due to the flat density profile, further inward measurements are not possible. In addition, the frequency interval \SIrange{140}{450}{\kilo\hertz} (green squares), deduced for the radial region with QC-mode activity, follows the velocity estimated for \SIrange{5}{450}{\kilo\hertz} with a small additional offset. The black circles denote the $E\times B$-velocity as it is estimated from the mean stationary profiles with the assumption of $Z_{eff}=\SI{1.5}{}~\cite{Wolf:2019}$. The neoclassical velocity is in agreement with the experimental determined velocities as long as the QC-mode becomes not visible and $T_e\approx T_i$, however, a small offset between the experimental- and $v_{E\times B}$-velocity is observed. Towards the plasma core, the deviation between experimental velocities and the  $v_{E\times B}$-velocity is increasing. This observation shows strong evidence that the QC-mode velocity is in electron diamagnetic direction, which is in favour for TEMs.\\The rotation profiles for both frequency intervals are estimated for each power step. For the frequency interval \SIrange{-110}{110}{\kilo\hertz}, including the $E\times B$-rotation, the profile is shown in fig.~\ref{fig:ExB-rotation}. Outside the minor radius of $a=$\SI{0.52}{\meter} the measured rotation is positive. After passing the velocity shear layer, the rotation is $v_\perp\approx\SI{-3}{\kilo\meter\per\second}$ and it increases towards the plasma core. This behaviour is the same for all power steps, indicating that the rotation is independent of the power level. Calculating the velocities for the frequency interval \SIrange{140}{450}{\kilo\hertz}, where the QC-mode activity is located, the rotation profile is different (see fig.~\ref{fig:QC-rotation}). Here, the estimation is possible for $r_{eff}\le\SI{0.4}{\meter}$, only. The reason becomes obvious when looking at the coherence spectrogram (fig.~\ref{fig:CoherenceSpectrum}), where the QC-mode activity starts at \SI{16.44}{\second}, corresponding to $r_{eff}\le\SI{0.4}{\meter}$. Furthermore, the QC-mode rotation shows a clear dependency on the power level, respectively the electron temperature gradient, ranging from \SI{-6}{\kilo\meter\per\second} for $P_{ECRH}=\SI{5.1}{\mega\watt}$ to \SI{-4}{\kilo\meter\per\second} for $P_{ECRH}=\SI{2.1}{\mega\watt}$. Comparing the QC-mode rotation with the low frequency turbulence rotation, including the $E\times B$-rotation, a factor $\approx 2$ faster QC-mode rotation is observed for each power step. It means the QC-mode rotates in electron diamagnetic drift direction, as it is expected for TEMs and for the universal instability, which is driven by the density gradient.\\Further evidence for a rotation in the electron diamagnetic drift direction is obtained when the rotation for each combination and the different frequency intervals is analysed (see fig.~\ref{fig:vperp-vs-distance}) averaged for the time interval with QC-mode activity. It shows that the rotation depends on the frequency range and the poloidal distance. Two facts are seen in fig.~\ref{fig:vperp-vs-distance}: (i) For the frequency interval \SIrange{-110}{110}{\kilo\hertz} the absolute velocities are smaller and decrease with poloidal distance, compared with the frequency intervals including the QC-mode. And (ii) the frequency intervals including the QC-mode yield within the error bars constant velocities, except the one for the smallest poloidal distance. From fig.~\ref{fig:Decomposition} it is seen that the QC-mode and LF-turbulence have a small overlap (red and green dashed Gaussian). Therefore, a small contribution of the QC-mode interval will contribute to the LF-turbulence interval. The contribution of the LF-turbulence decreases faster than the QC-mode and the contribution of the QC-mode to the LF-turbulence becomes stronger, yielding an increase in the absolute velocity. It indicates a faster decay of the low frequency turbulence. In addition, fig.~\ref{fig:vperp-vs-distance} supports that the rotation of the QC-mode is in electron diamagnetic drift direction for all poloidal distances.

\subsection{QC-mode properties}\label{QC-Properties}
\begin{figure}
    \centering
    \includegraphics[scale=0.8]{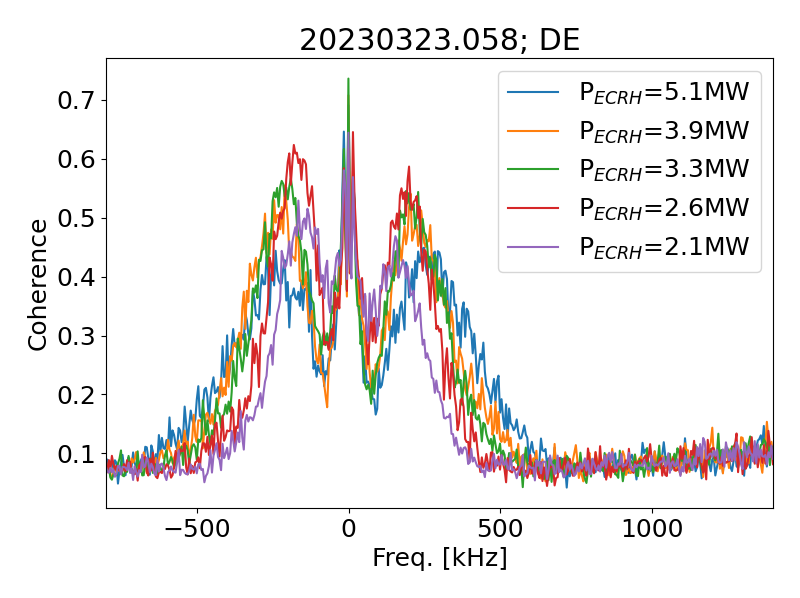}
    \caption{Coherence spectra for 5 different power levels for program \textit{20230323.058}, showing the decrease of $f_{QC}$ with power. The spectra are obtained for the antenna combination DE probing a poloidal distance of $\Delta z = \SI{17}{\milli\meter}$.}
    \label{fig:PowerDependence}
\end{figure}

For each power step, the coherence spectrum is calculated for an antenna combination with a medium poloidal distance of \SI{17}{\milli\meter} and where the reflectometry scan was not disturbed by the modulation of the ECRH power. All coherence spectra see fig~\ref{fig:PowerDependence} show the same broad structure, however, the centre mode frequency ($f_{QC}$) and coherence of the QC-mode are varying with the power steps. While $f_{QC}$ and the FWHM of the QC-mode are decreasing with the power, indicating a dependence on $\nabla T_e$, the coherence has a maximum at medium power levels of \SI{2.6}{\mega\watt}$\le P_{ECRH}\le$\SI{3.3}{\mega\watt}.

\begin{figure}
    \centering
    \includegraphics[scale=1]{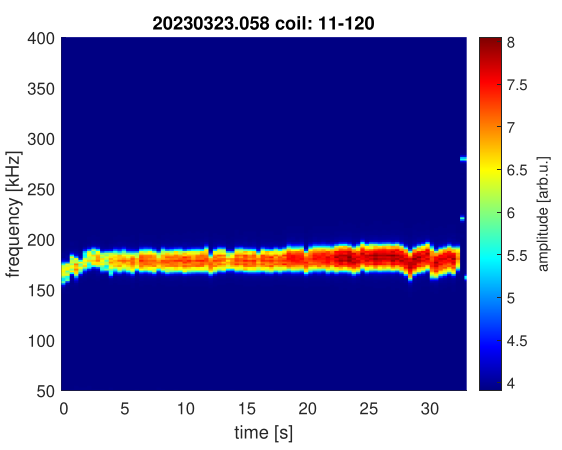}
    \caption{Mode structure in the Mirnov coil for \textit{20230323.058} after DMUSIC processing. The mode structure shows no dependence of the power and for $t\le\SI{3}{\second}$ the mode frequency decreases.}
    \label{fig:Mirnov}
\end{figure}

\noindent To exclude that the observed QC-mode has a magnetic component, the coherence spectrum is compared with data taken by Mirnov coils. The Mirnov coils are most sensitive for the low-mode-number MHD- and Alfv\'en-modes in W7-X. In case, the QC-mode has a large poloidal extension which is similar to a low poloidal mode number, it should be visible in the Mirnov signals as well. Using the DMUSIC~\cite{schmidt_multiple_1986} algorithm, Mirnov coils show indeed a broad-band structure ranging from \SIrange{160}{190}{\kilo\hertz} whose amplitude increases with decreasing power level. As seen from fig.~\ref{fig:Mirnov}, for the Mirnov coil QXM11CE120x	installed at the poloidal angle $\theta=\SI{9.48}{\degree}$, the amplitude is increasing while the centre frequency of the structure observed in the Mirnov coils is constant for all power levels. Only within the first \SI{3}{\second} of the program, an increasing frequency is observed with decreasing density, indicating an Alfv\'enic nature~\cite{Rahbarnia:2021, VazMendez:2023}. However, a constant frequency, independent of the applied heating power $P_{ECRH}$ is not observed by PCR, and indicates the absence of a magnetic component in the QC-mode observed by PCR.

\begin{figure}
    \centering
    \includegraphics[scale=0.55]{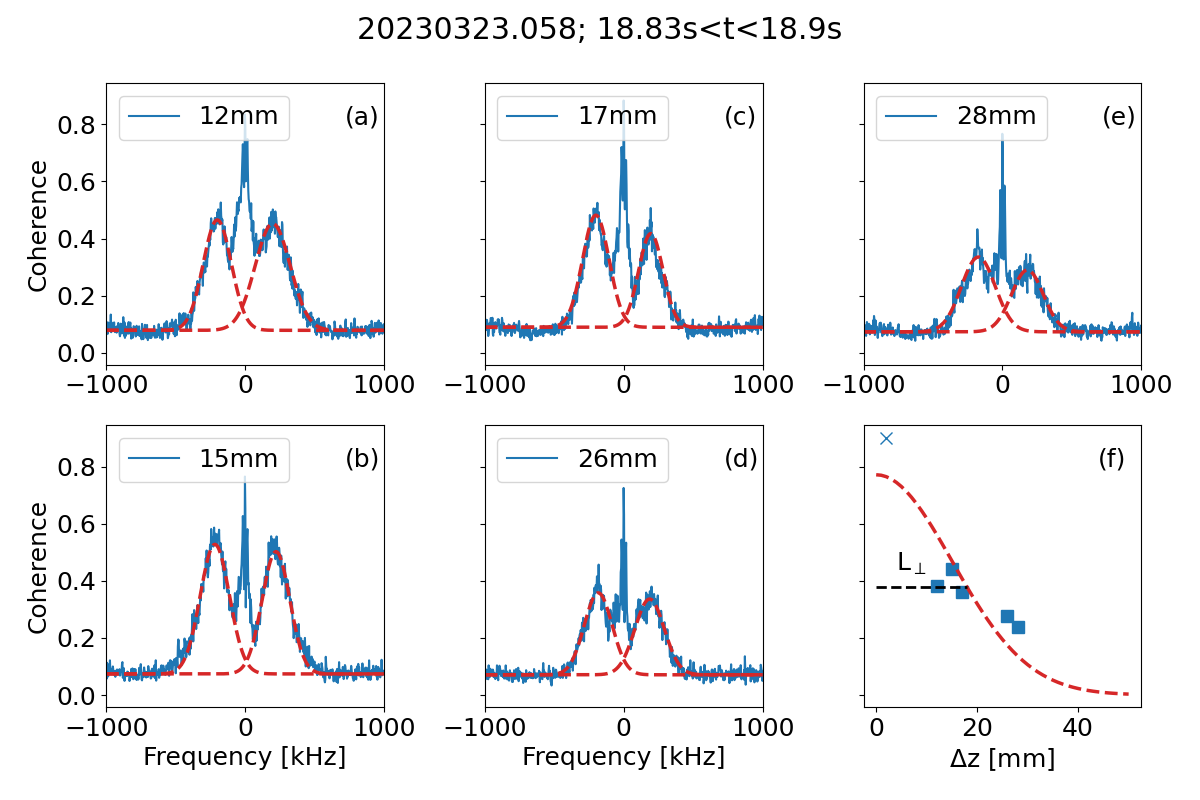}
    \caption{Coherence spectra for the standard configuration and for different poloidal distances as indicated in the label, showing the decrease of the QC-mode coherence (a-e). In addition, the red dashed line shows the fit calculated QC-mode component for both frequency branches. The poloidal correlation of the QC-mode is shown in (f) yielding a HWHM structure length of $L_\perp=$\SI{18}{\milli\meter}}
    \label{fig:Coherence4z_EIM}
\end{figure}

\noindent To get an estimate of the poloidal size of the QC-mode, the coherence is calculated for all 6 antenna combinations. The calculation is performed for the time interval with QC-mode activity. For all combinations, the resulting coherence spectrum is analysed by decomposition in a low frequency component and QC-mode components, as discussed in section~\ref{QC-Velocity}. As long as the spectra contain these 3 components, only, the measured coherence can be used for the determination of the poloidal size of the QC-mode. For the smallest poloidal distance a large contribution of broad-band turbulence disturbs the measurement of the QC-mode amplitude, therefore it is not used in the analysis. The coherence spectra for the power step of $P_{ECRH}=$\SI{3.3}{\mega\watt} are presented in fig.~\ref{fig:Coherence4z_EIM}(a-e), where the coherence spectra are shown for a \SI{12}{\milli\meter}$\le\Delta z\le$\SI{28}{mm}. The coherence spectra at the shortest distance cannot be used because the spectra is dominated by fast decaying broad band turbulence. The given $\Delta z$ values are corrected for the field line pitch, which is calculated from the ratio of the $\Delta t$-values~\cite{akfl:2010} obtained from the cross correlation analysis. The poloidal distance is estimated as a mean value within the radial range where the QC-mode is observed. To show the good agreement of the QC-mode components, described by two Gaussian shapes, with the measured spectrum, the QC-mode contribution is highlighted by red dashed lines. In fig.~\ref{fig:Coherence4z_EIM}f the QC-mode coherence is shown as a function of the poloidal distance probed by the different antenna combinations. For the reason outlined above, the coherence at $\Delta z=\SI{2}{\milli\meter}$ is set to $\gamma\approx 0.9$, indicated by a $\times$-symbol. The poloidal decay of the coherence is again approximated by a Gaussian. The estimated half width at half maximum (HWHM) of the poloidal structure length yields $L_\perp=$\SI{18}{\milli\meter}. Recalling the poloidal size of TEMs which is in the order of \SIrange{20}{30}{\milli\meter}~\cite{Doyle:2007, Edlund:2018}, the experimental obtained $L_\perp$ is in agreement with the assumption of TEMs.

\begin{figure}
    \centering
    \includegraphics[scale=0.7]{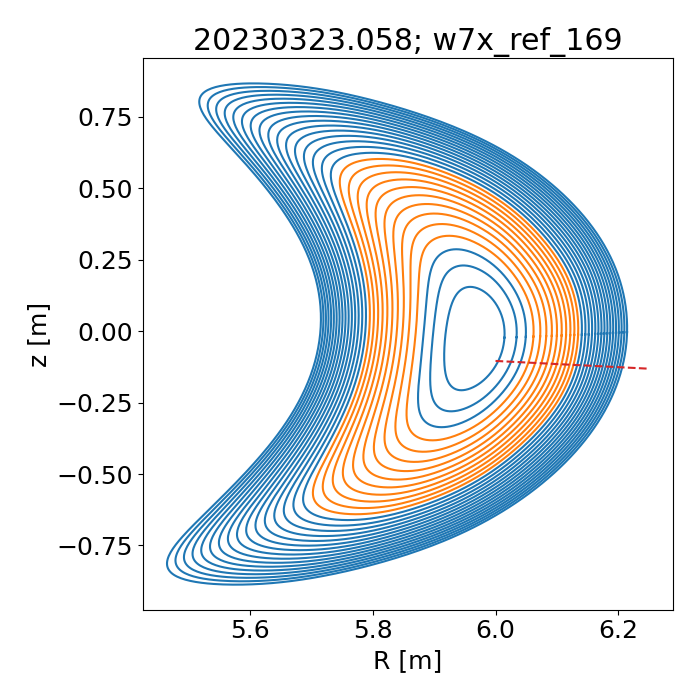}
    \caption{Flux-surface in a poloidal plane calculated by VMEC equilibrium for the program \textit{20230323.058}. The orange flux surfaces denote the radial range with QC-mode activity. The red dashed line is the line of sight of the PCR launching antenna}
    \label{fig:VMEC}
\end{figure}

\noindent For an independent proof of the poloidal structure length of the QC-mode, the flux surface geometry at the toroidal position of the PCR is calculated from the VMEC-equilibrium. The radial range of the QC-mode activity is highlighted in orange colour (see fig.~\ref{fig:VMEC}). Having estimated the mode rotation and mode frequency, a relation between the poloidal size of the QC-mode, the rotation and the mode frequency can be applied \begin{equation}\label{equ:mode-number}
    m = \frac{s\, f_{QC}}{v_{QC}},
\end{equation} where $m$ denotes the mode number, which is related to the poloidal size of the mode, and $s$ is the poloidal circumference of the flux surface where the QC-mode is observed. The calculated circumference varies from \SIrange{1.57}{2.74}{\meter}. Due to the fact that the mode velocity and the frequency are nearly constant, the mode number varies radially between \SIrange{80}{120}{}, keeping the poloidal structure size at $s/m = L_\perp\approx$ \SI{21}{\milli\meter} for one power step. For all power steps, a mean mode number of $\overline{m}=\SI{101}{}\pm\SI{5}{}$ and a mean poloidal structure length of $\overline{L}_\perp=\SI{21}{\milli\meter}\pm\SI{1}{\milli\meter}$ is estimated. The resulting mean wave number is calculated as $\overline{k}_\perp=\SI{2.95}{\per\centi\meter}\pm\SI{0.14}{\per\centi\meter}$. This value agrees well with the one that is obtained from the decay of the coherence as function of the poloidal distance of the antenna, thus, the two independent methods to extract the poloidal size of the QC-mode structure yield the same result. Moreover, the latter method estimates a mean $L_\perp$- and $k_\perp$-value, respectively. The agreement with the $L_\perp$- and $k_\perp$-value calculated from the poloidal position of the PCR indicates that a possible  poloidal variation of these values seems to be negligible on the investigated flux surfaces. To calculate the ion sound radius ($\rho_s$), the electron temperature for each power step is estimated at the mean radius of $r_{eff}=\SI{0.28}{\meter}$. Due to the decreasing power steps as function of time, the electron temperature decreases from $T_e=\SI{1480}{eV}$ to $T_e=\SI{870}{eV}$ and the ion temperature decreases from $T_i=\SI{1130}{eV}$ to $T_i=\SI{880}{eV}$. The local magnetic field at this position amounts to $B_t=\SI{2.31}{\tesla}$, so that $1.22\leq k_\perp\rho_s\le1.51$ is calculated from the experiment.

\begin{figure}
    \centering
    \includegraphics[scale=0.8]{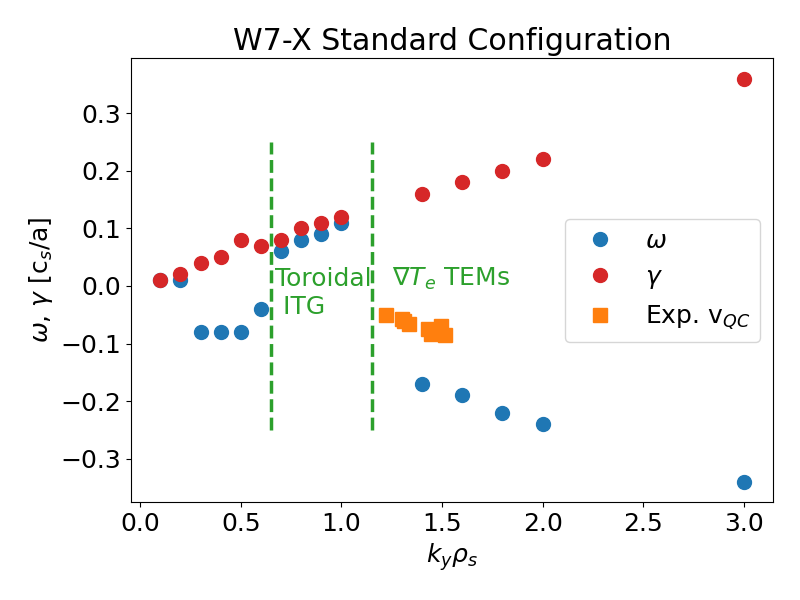}
    \caption{Growth rate and angular frequency of the mode in units of $c_s/a$ as function of $k_y\rho_s$ for the parameters of program \textit{20230323.058}. The transition from ITG to TEM takes place at $k_y\rho_s\ge 1$. The interval of measured $k_\perp\rho_s$-values is indicated by the horizontal bar.The squares denote the $f_{QC}$ estimated from the experimental measured velocities}
    \label{fig:growth-rate}
\end{figure}

\noindent To prove the assumption that the observed QC-mode has a TEM nature, linear gyrokinetic simulations using the \textsc{GENE} code \cite{Jenko2000} are performed. The calculations use the plasma parameters of the program \textit{20230323.058} and are performed in the so-called bean flux tube of W7-X. The results in fig.~\ref{fig:growth-rate} show the growth rate ($\gamma$) and the angular rotation ($\omega$) of the fastest-growing mode at a given poloidal wave number $k_y\rho_s$. The mode velocity is calculated in the $E\times B$ frame. Two regions, separated by green lines, can be identified. For $0.65\le k_y\rho_s\le 1.15$, $\omega$ is positive and this region is related to toroidal ITG-modes (propagation in ion diamagnetic drift direction). For $k_y\rho_s\ge 1.15$, $\omega$ becomes negative and this region is related to $\nabla T_e$-driven TEMs, (propagation in electron diamagnetic drift direction). The transition takes place for $1.0 \le k_y\rho_s\le 1.35$. The experimental $k_\perp\rho_s$-values cover partly the region where the transition from ITG-modes to $\nabla T_e$ driven TEMs is expected according to the simulations outlined above. However, most of the experimentally deduced $k_\perp\rho_s$-values are located in the region of $\nabla T_e$ driven TEMs.\\To compare the angular velocity from the simulations  with the experimental values, the ion sound speed $c_s$ is calculated for the mean radial position $r_{eff}$, where the QC-modes are observed, for each power step. The minor radius for the program is $a=\SI{0.51}{\meter}$. From fig.~\ref{fig:growth-rate} the range of $\omega$ in units of $c_s/a$ is \SIrange{-0.15}{-0.18}{\kilo\hertz}. Using the measured velocities from the experiment, the frequency of the mode in the $E\times B$-frame and in units of $c_s/a$ is estimated as:\begin{equation}
    \omega_{QC} = \frac{2\pi\,a\,(v_\perp^{QC}-v_{E\times B})}{r_{eff}\,c_s}
\end{equation} Here, $v_\perp^{QC}$ denotes the velocity for the QC-mode. The $E\times B$-velocity is taken from (i) the neoclassical calculations or (ii) $v_\perp^{LF}$, the velocity of the low frequency turbulence as estimated from PCR. For both data sets, the resulting $\omega$ in units of $c_s/a$ for the different power steps decreases with increasing $k_\perp\rho_s$. Taking the data from neoclassical calculations $\omega_{QC}$ is in the range \SIrange{-0.05}{-0.08}{\kilo\hertz} is achieved which is a factor of $\le3$ smaller than expected $\omega_{QC}$ from the linear numerical simulations. However, nonlinear gyrokinetic calculations performed for HSX~\cite{faber:2015} have shown a broad range of TEM-mode frequencies, even below those values obtained from linear gyrokinetic calculations. These are most likely to be attributed to the large plethora of subdominant modes, which are frequently found in low-shear devices such as HSX and W7-X, and which can inhabit a large range of frequencies~\cite{Pueschel:2016}, similar to those found in the experiments on W7-X.\\
Taking all the measurements together (i) rotation in the electron diamagnetic drift direction, (ii) the poloidal size of $L_\perp=\SI{21}{\milli\meter}$ and  (iii) the range of $k_\perp\rho_s$ of $1.22\leq k_\perp\rho_s\le1.51$ strong evidence is found that the QC-modes in the experiment are caused by $\nabla T_e$-driven TEMs.

\subsection{Effects of the magnetic configuration}\label{MagnConf}
\noindent Up to here, we investigated the standard magnetic configuration (EIM), where indications of TEMs in the plasma core were found. In general, TEMs require a magnetic mirror to exist, and typically its growth rate and severity are proportional to the fraction of trapped particles. The mirror term in W7-X can be varied by changing the current ratios in the non-planar coils. If a change in the mirror ratio changes the QC-mode observation, it is additional evidence for the existence of TEMs in W7-X. Two additional configurations are thus investigated: the low-mirror configuration (AIM) with a mirror ratio of \SI{5.5}{\percent} and the high-mirror configuration (KKM) with a mirror ratio of \SI{8.8}{\percent}. For both configurations, a similar program with power steps is used (see fig.~\ref{fig:Overview}). The analysis is the same as outlined before. 

\begin{figure}
    \centering
    \includegraphics[scale=0.6]{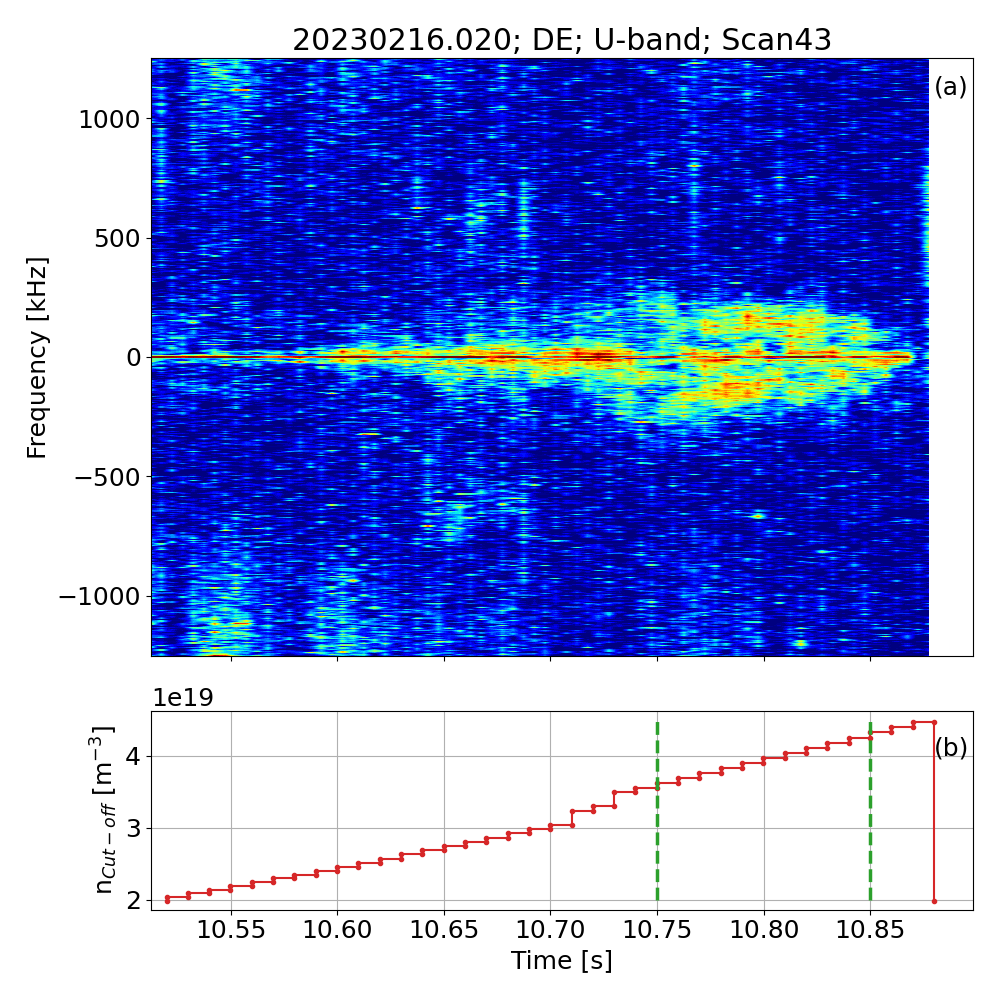}
    \caption{Coherence spectrogram (a) for one scan of the PCR diagnostic, showing the broad quasi QC-mode structure and the related electron density for a program in low-mirror configuration. The coherence spectrum is calculated for combination DE with $\Delta z = \SI{17}{\milli\meter}$. (b) of the reflectometer scan. The green dashed lines denote the density range where the QC-mode is observed.}
    \label{fig:QC-mode4low-mirror}
\end{figure}

\begin{figure}
    \centering
    \includegraphics[scale=0.55]{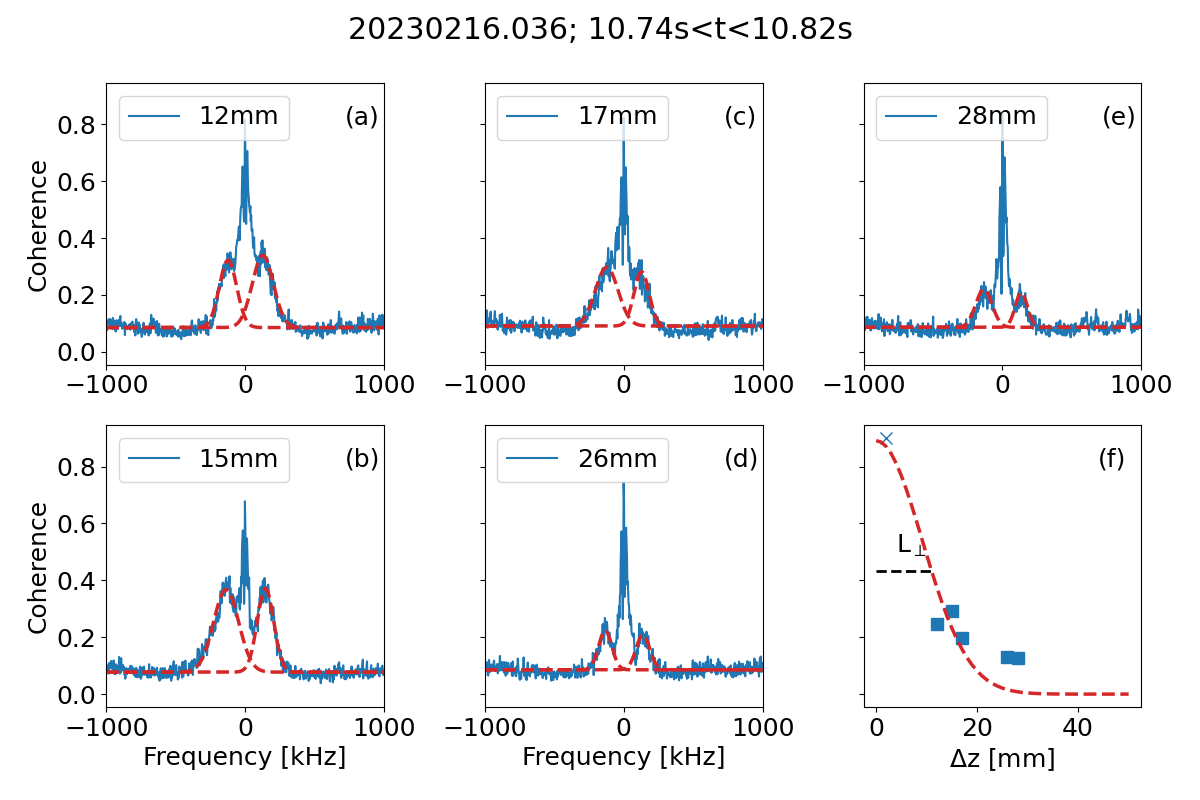}
    \caption{Coherence spectra for the low mirror configuration and for different poloidal distances as indicated in the label showing the decrease of the QC-mode coherence (a-e). In addition, the red dashed line shows the fit calculated QC-mode component for both frequency branches. The poloidal correlation of the QC-mode is shown in (f) yielding a HWHM structure length of $L_\perp=$\SI{14}{\milli\meter}}
    \label{fig:Coherence4z_AIM}
\end{figure}

\noindent The analysis is repeated for the low-mirror configuration and for the program \textit{20230216.020, .036}. The coherence spectrogram for a power of $P_{ECRH}=$\SI{3.23}{\mega\watt} is shown in fig.~\ref{fig:QC-mode4low-mirror}. A large difference in the centre frequency of the QC-mode is found ($f_{QC}=$\SI{132}{\kilo\hertz}), when compared with the QC-mode in program in standard configuration, where $f_{QC}=$\SI{196}{\kilo\hertz} was observed. Assuming a TEM origin, the difference in the observed QC-mode frequencies could be (i) a change in electron-temperature scale length $L_{Te}$ and/or (ii) a difference in the rotation of the mode. From $T_e$-profiles from Thomson Scattering, the estimate of the scale length at a similar power level yield $L_{Te}=$\SI{0.164}{\meter} for both discharges. For the programs \textit{20230216.020, 20230216.036} the estimated poloidal structure length, estimated from all antenna combination and excluding the shortest one, amounts to $L_\perp=$\SIrange{12}{14}{\milli\meter} (see fig.~\ref{fig:Coherence4z_AIM}). These values are in agreement with the theoretically expected wave length of TEMs and supports the notion that the QC-modes are of TEM nature. 

\begin{figure}
    \centering
    \includegraphics[scale=0.8]{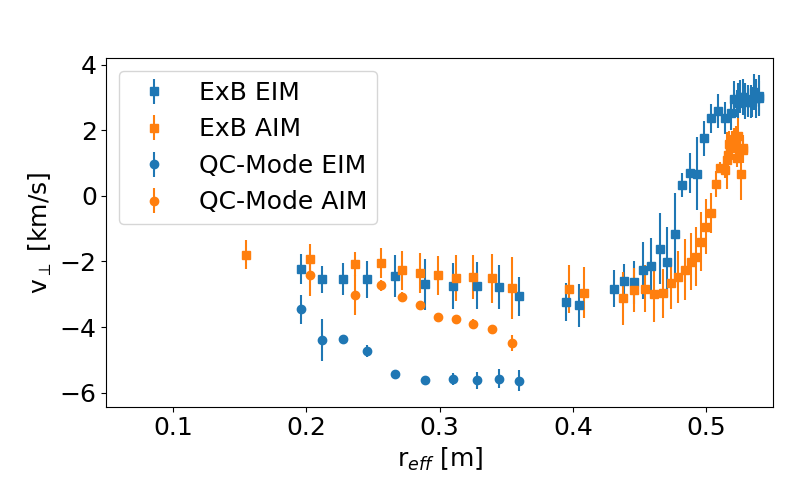}
    \caption{QC-mode rotation (circle markers) and $E\times B$ rotation (square markers) for the programs \textit{20230323.058} (EIM) and \textit{20230216.020} (AIM)for a time interval of \SI{6.14}{\second}$\le t\le$\SI{6.6}{\second}. Whereas the $E\times B$ rotation shows no difference for both configurations, the QC-mode rotation is largest for the program \textit{20230323.058} (blue circles). In both cases, the QC-mode rotation is faster than the $E\times B$ rotation.}
    \label{fig:ExB-and-QC-mode}
\end{figure}

\noindent For the low-mirror configuration, the decomposition of the coherence spectra is more complicated, because the QC-mode Gaussian has an overlap with the low-frequency turbulence Gaussian, including the $E\times B$-rotation. In this case, the frequency range is set to \SIrange{5}{80}{\kilo\hertz} for estimating the LF-turbulence-rotation including the $E\times B$-rotation and \SIrange{80}{350}{\kilo\hertz} for the estimation of the QC-mode rotation. The rotation for both frequency intervals is compared for the programs \textit{20230323.058, EIM} and \textit{20230216.020, AIM} for a time interval of \SI{6.14}{\second}$\le t\le$\SI{6.6}{\second}, where P$_{ECRH}=\SI{4.3}{\mega\watt}$. In fig.~\ref{fig:ExB-and-QC-mode} the rotation profiles for both programs are shown. While the $E\times B$-rotation velocities with $v_{E\times B}=\SI{-2.5}{\kilo\meter\per\second}$ are equal within the error bars and shows a slight increase towards the plasma core, the measurement of the QC-mode rotation shows large differences. For the EIM configuration, a rotation speed of $v_{QC}=$\SI{-5.5}{\kilo\meter\per\second} is estimated, which increases for $r_{eff}\le$\SI{0.25}{\meter} to \SI{-3.5}{\kilo\meter\per\second}. The QC-mode rotation in AIM shows an increase with decreasing $r_{eff}$ from \SIrange{-4.0}{-2.5}{\kilo\meter\per\second} and is comparable within the error bars with the $E\times B$-rotation at its innermost radius. The difference in the rotation seems to be related to the higher frequency of the QC-mode in the EIM-configuration (program \textit{20230323.058}). It amounts to a factor of $\approx$\SI{1.6}{} as estimated from the radius averaged coherence spectra and suggests a relation between the QC-mode frequency and the QC-mode rotation. Concluding these observations, an agreement with a TEM-nature of the QC-mode is evident, because in both cases the QC-mode rotation is in electron diamagnetic drift direction for almost the full radial observation interval of the mode.

\begin{figure}
    \centering
    \includegraphics[scale=0.7]{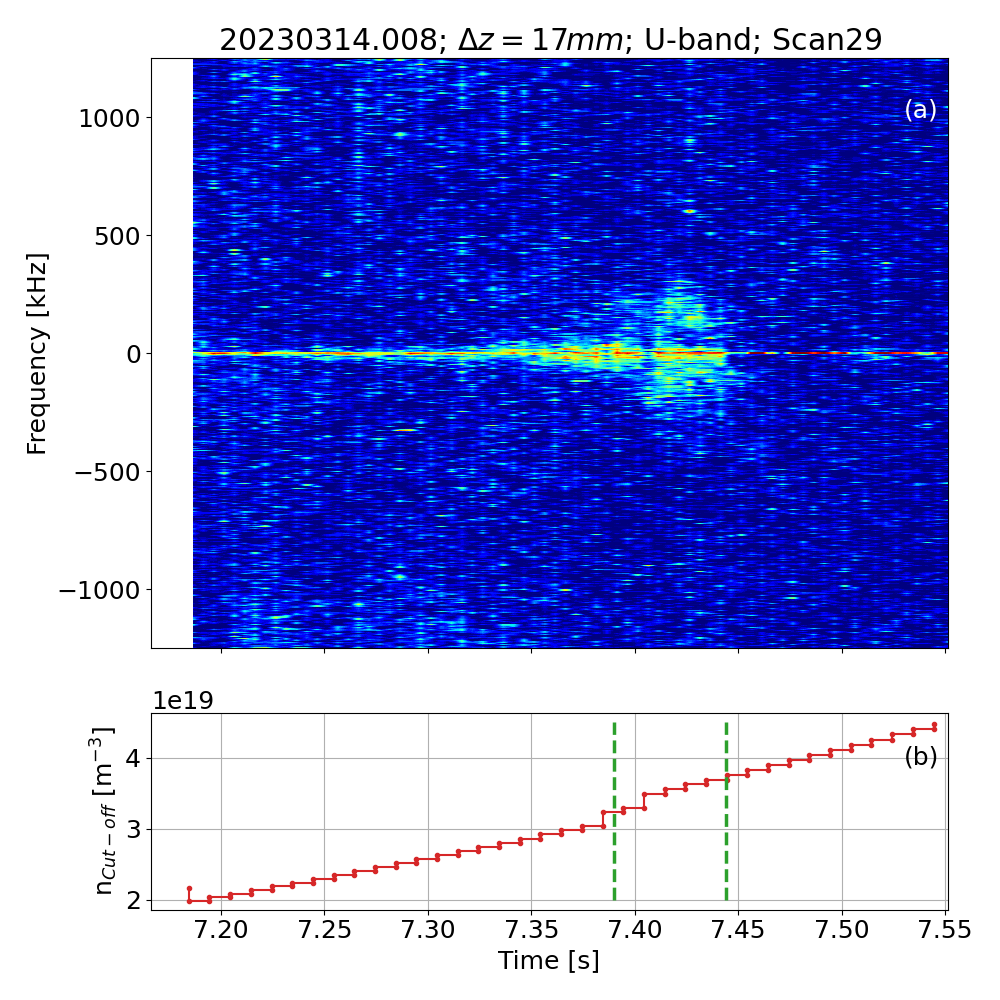}
    \caption{Coherence spectrogram (a) for high mirror configuration for an antenna combination with $\Delta z=\SI{17}{\milli\meter}$. It shows the existence of a weak QC-mode in the plasma core, (b) shows the density staircase of this scan. The time interval with QC-mode activity is marked by dashed green lines.}
    \label{fig:CoherenceKKM}
\end{figure}

\begin{figure}
    \centering
    \includegraphics[scale=0.7]{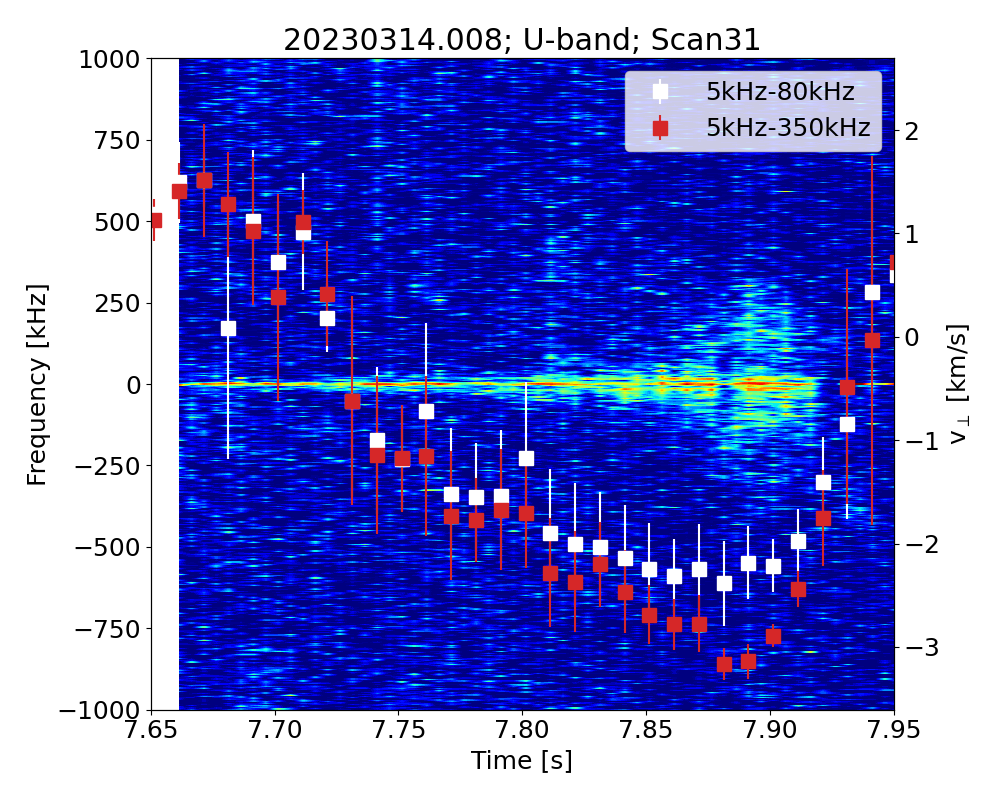}
    \caption{Coherence spectrogram for scan 31 with overlaid velocity as estimated from all antenna combination, except the smallest one. The low frequency range (white squares) describes the $E\times B$-rotation and the high frequency interval (red squares) the $E\times B$ plus QC-mode rotation. A clear increase in electron diamagnetic drift direction is observed for the time interval where the mode is present. Furthermore, the deviation is largest when the mode becomes strongest at $t=\SIrange{7.87}{7.91}{\second}$.}
    \label{fig:Coherence4Delays_KKM}
\end{figure}

\begin{figure}
    \centering
    \includegraphics[scale=0.8]{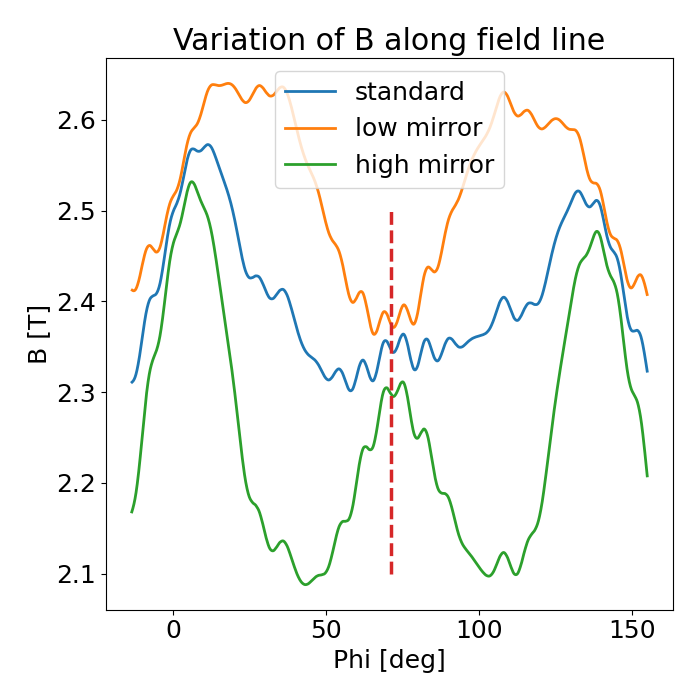}
    \caption{Magnetic field of a field line intersecting the LoS of the PCR (red dashed line) in the radial range where the QC-modes are observed. It shows the existence of a magnetic mirror in all three cases. However, for the high mirror case the PCR is not located in the minimum of $B$ which explains qualitatively the weak QC-mode observation at the PCR position.}
    \label{fig:MirrorPresentation}
\end{figure}

\noindent The last investigated configuration is the high-mirror configuration (KKM). As shown in \ref{fig:Overview}, the program \textit{20230314.008} is selected for further analysis. It has a line averaged density of \SI{5.2e19}{\per\square\meter} and a power scan ramping down from \SI{4.3}{\mega\watt} to \SI{2.3}{\mega\watt}. For the scan ranging from \SIrange{7.2}{7.55}{\second}, a weak QC-mode is observed as shown in fig~\ref{fig:CoherenceKKM} with comparable centre frequencies as in the low-mirror configuration (AIM). The QC-mode is observed for all power steps in the program, however, the time interval for the mode observation is decreased from $\approx\SI{100}{\milli\second}$ in the low mirror case to $\approx\SI{40}{\milli\second}$ in the high mirror case. As in the AIM cases, the propagation time is calculated for two frequency intervals: (i) \SIrange{-5}{-80}{\kilo\hertz} and \SIrange{5}{80}{\kilo\hertz}, respectively, determining the $E\times B$-velocity and (ii) \SIrange{-5}{-350}{\kilo\hertz} and \SIrange{5}{350}{\kilo\hertz} covering QC-mode and the $E\times B$ velocity. In a further step, a mean $v_\perp$ is estimated from all antenna combinations, except the combination probing the shortest poloidal distance. For the time interval \SIrange{7.5}{8.0}{\second}, the coherence spectrogram is shown in fig.~\ref{fig:Coherence4Delays_KKM} and overlaid with $v_\perp$ as determined from the elliptical model, for the antenna combination with a poloidal separation of $\Delta z=\SI{17}{\milli\meter}$. While up to \SI{7.8}{\second} the difference between $v_\perp$ for the two frequency intervals is equal within the error bars, they start to deviate with the onset of the QC-mode at $t=\SI{7.87}{\second}$. At the maximum of the QC-mode activity at \SI{7.9}{\second}, the sum of QC-mode and $E\times B$-rotation exceeds the $E\times B$-rotation by a factor of \SI{1.4}{}. This indicates that also in the high-mirror configuration, the observed QC-modes are likely caused by TEMs.\\To understand the weak amplitude of the QC-modes in the KKM configuration, the toroidal variation of the magnetic field along a field line which includes the probing position of the PCR during QC-mode activity, is investigated. In fig.~\ref{fig:MirrorPresentation} the local magnetic field as function of the toroidal angle is shown. The dashed lined indicates the probing position of the PCR diagnostic. It is clearly seen that the probing position of the PCR during QC-mode activity in the standard and low mirror configuration is located in the minimum of the magnetic field. However, for the high-mirror case, the probing position of the PCR diagnostic is located in a local maximum of magnetic field. This explains qualitatively the weak QC-mode amplitude in the high-mirror configuration. A further result drawn from this figure is the possibility of QC-modes for diagnostics located at a different toroidal angle, where the minimum in the magnetic field is localized. 

\subsection{Effect of NBI-heating}
\begin{figure}
    \centering
    \includegraphics[scale=0.6]{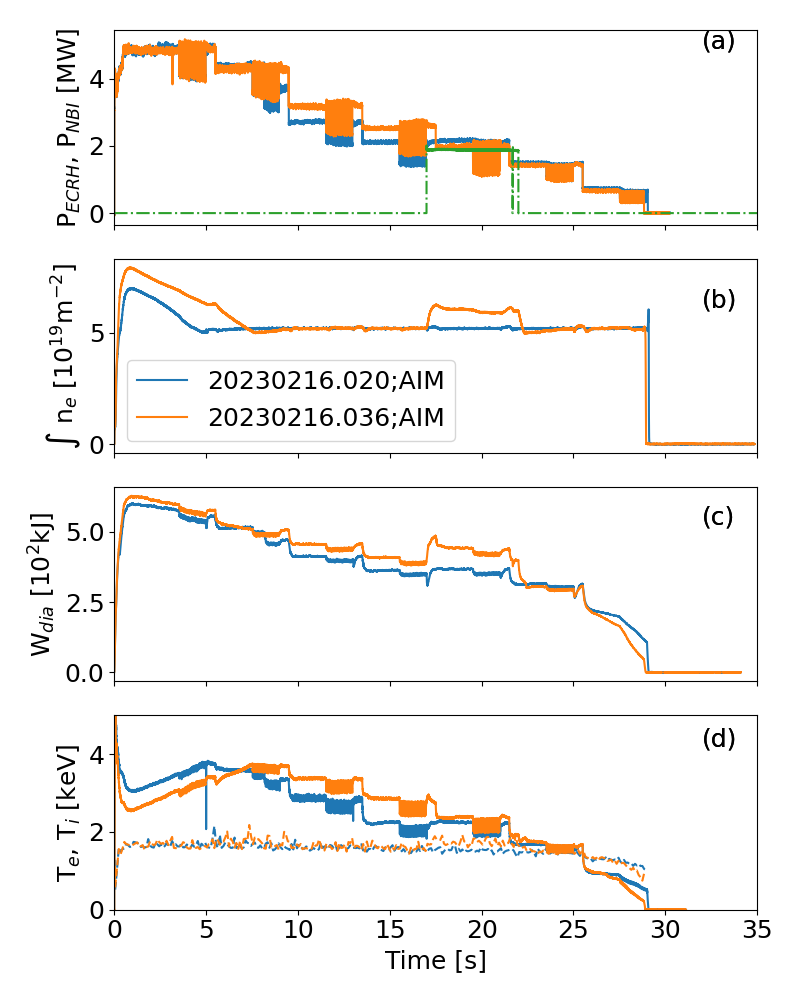}
    \caption{The important parameters of the investigated programs \textit{20230216.020} and \textit{20230216.036} with NBI-injection as function of time. (a) ECRH-power and NBI-power, (b) line-averaged density, (c) diamagnetic energy and (d) central electron (solid) and ion (dashed) temperature. Clearly seen is the increase of density and diamagnetic energy during the NBI-phase.}
    \label{fig:Overview230216020_036}
\end{figure}

\noindent Until now, only programs with ECRH-power have been investigated. The program \textit{20230216.036}, in low-mirror configuration, offers the possibility to investigate a time interval where neutral beam heating is applied, too. In this interval the applied heating amounts to $P_{tot}=\SI{3.8}{\mega\watt}$, with $P_{ECRH}=$\SI{2.0}{\mega\watt} and $P_{NBI}=\SI{1.8}{\mega\watt}$. The beam fuelling of the NBI increases the line-averaged density from \SI{5e19}{\per\square\meter} to \SI{6e19}{\per\square\meter}. A comparison regarding the main plasma parameters of both programs is shown in fig.~\ref{fig:Overview230216020_036}. Beside the similar ECRH-heating pattern, the dashed-dotted line shows the NBI-injection for program \textit{20230216.036}. Clearly seen is the increase of the line-averaged density and diamagnetic energy for the NBI-phase. Also, a small increase in the ion temperature can be recognized. The analysis for this time interval yields a centre frequency of the QC-mode of $f_c=\SI{149}{\kilo\hertz}\pm\SI{13}{\kilo\hertz}$. The velocity for the same interval and in the frequency range \SIrange{80}{350}{\kilo\hertz} is estimated to be $v_\perp=\SI{-3.9}{\kilo\meter\per\second}\pm\SI{0.2}{\kilo\meter\per\second}$. This is within error bars close to the value which is obtained for a power step of $P_{ECRH}=\SI{4.3}{\mega\watt}$ for the same configuration in programs \textit{20230216.020} with $f_c=\SI{148}{\kilo\hertz}$ and $v_\perp=\SI{-3.5}{\kilo\meter\per\second}\pm\SI{0.6}{\kilo\meter\per\second}$ as well as for program \textit{20230216.036}, yielding $f_c=\SI{161}{\kilo\hertz}$ and $v_\perp=\SI{-3.8}{\kilo\meter\per\second}\pm\SI{0.5}{\kilo\meter\per\second}$. However, the $\nabla T_e$ is different for the phase with NBI-injection. It demonstrates that beside $\nabla T_e$ also other quantities as $\nabla n_e$ or collisionality could play an important role in the driving mechanism for the QC-modes. It needs further studies, which are outside the scope of this paper, to analyse and understand this observation.

\section{Relation to Tokamak Observations}\label{Rel}
\noindent The results from W7-X support a general scaling of the QC-mode frequency with the mode rotation. To support this hypothesis, observations from other devices are presented. The comparison with tokamak data helps to validate the universality of the QC-mode scaling observed in W7-X, potentially extending insights into mode behaviour across different magnetic confinement geometries. Here, two examples from limiter tokamaks are presented where QC-mode studies have been performed in the last years. One device is Tore Supra (TS) and the second one is TEXTOR. In both devices, QC-modes are observed by an X-mode hopping frequency reflectometer and Poloidal Correlation Reflectometry, respectively. The next two subsections summarize the observations at both devices and demonstrate that the scaling found for W7-X is suitable to describe the observations at TS and TEXTOR.
\subsection{QC-Modes at Tore Supra}
\noindent Experiments at the circular limiter tokamak Tore Supra with a major radius of $R_0=$\SI{2.4}{\meter} and a minor radius of $a=$\SI{0.72}{\meter} are performed to study QC-modes in the plasma core~\cite{arnichand_quasi-coherent_2014}. These modes are observed during the linear ohmic confinement regime (LOC) and it is found that they disappear if the density is increased and the confinement is saturated, the so-called saturated ohmic confinement regime (SOC). The QC-modes are observed by using an X-mode hopping-frequency reflectometer~\cite{Clairet:2010}. A detailed analysis based on experimental observations, gyrokinetic simulations and synthetic diagnostic leads to the interpretation that the QC-mode have a TEM nature. To prove that the QC-modes fulfil the W7-X scaling, the QC-mode frequency of $fc=$\SI{60}{\kilo\hertz} from ref.\cite{arnichand_identification_2016} is taken as a starting point. The mode is detected at $r/a=$\SI{0.18}{} and yields a radius of the circular flux surface of \SI{0.13}{\meter}. Assuming that the poloidal structure of the QC-mode amounts to $L_{QC}=$\SI{25}{\milli\meter} the poloidal mode number is calculated to be $m\approx$\SI{33}{}. This allows to calculate the QC-mode velocity, which yields $v_{QC}$\SI{-1.5}{\kilo\meter\per\second}. The investigated plasmas at Tore Supra show sawtooth activity in the plasma core. This activity is accompanied by $m/n=1/1$ sawtooth crash precursor. From the measurement of the sawtooth precursor frequency, an estimate of the $E\times B$-rotation can be deduced. The precursor frequency of $f_{pc}\approx$\SI{1.3}{\kilo\hertz}~\cite{Salazar:2023} yields a $E\times B$-rotation of $\approx$\SI{-1}{\kilo\meter\per\second} and confirms that the QC-mode rotation is faster than the $E\times B$-rotation and that it propagates in the electron diamagnetic drift direction. The Tore Supra data, due to the low QC-mode frequency, extends the measurement to the end of small $E\times B$-rotation and fits well to the scaling deduced from the W7-X data. Moreover, since the measurements at Tore Supra confirms the QC-modes as TEMs, this supports also that the QC-mode origin at W7-X is of TEM nature. 
\subsection{QC-Modes at TEXTOR}
\begin{figure}
    \centering
    \includegraphics[scale=0.6]{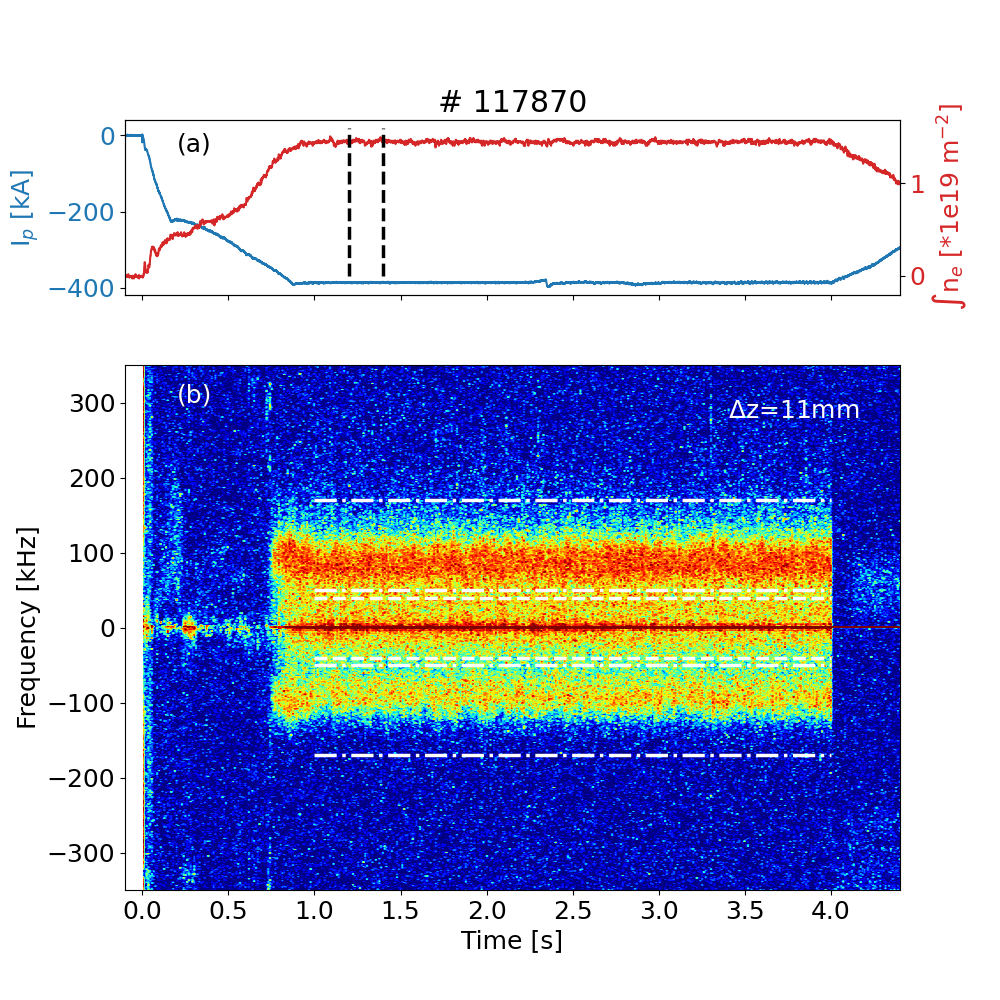}
    \caption{\textbf{(a)} Time traces of plasma current and line averaged density. The vertical dashed lines denote the time interval used for the analysis of $v_{QC}$. \textbf{(b)} Coherence spectrogram for fixed combination showing the QC-mode. The horizontal dashed line denote the frequency range used for estimation of the $E\times B$-rotation and the dashed dotted line denote the interval used for the QC-mode rotation.}
    \label{fig:TEXTOR-117870}
\end{figure}

\noindent The limiter tokamak TEXTOR~\cite{Neubauer:2005} with a major radius of $R_0=$\SI{1.75}{\meter} and a minor radius of $a=$\SI{0.5}{\meter} was equipped with a PCR system~\cite{akfl:2004} measuring at the low field side (LFS) and consisting of one launching and four receiving antennae. The antenna arrangement is similar compared to the one at W7-X. For the ohmic discharge \#117870 with a line averaged density of $\int n_e=$\SI{1.5e19}{\per\square\meter}, a plasma current of $I_p=$\SI{-400}{\kilo\ampere} and a toroidal magnetic field of $B_t=$\SI{1.9}{\tesla} the PCR was operating at a frequency of $f_{ref}=$\SI{35.3}{\giga\hertz}, which corresponds to a local electron density of $n_e=$\SI{1.55}{\per\cubic\meter}. After current ramp up, the discharge exhibits a long flat top ranging from \SIrange{0.9}{4}{\second}. This plasma shows strong QC-mode activity in the plasma core, as shown in fig.~\ref{fig:TEXTOR-117870} during the whole flat top phase. The analysis is performed within the time interval of \SIrange{1.2}{1.4}{\second} as indicated in fig.~\ref{fig:TEXTOR-117870}a. The centre frequency is observed at $f_{QC}=$\SI{85}{\kilo\hertz}. Applying the similar analysis as outlined above for W7-X, the coherence spectrum can be decomposed into a low-frequency turbulence part including the $E\times B$-rotation ranging from \SIrange{-40}{40}{\kilo\hertz} and the QC-mode component from \SIrange{-170}{-50}{\kilo\hertz} and \SIrange{50}{170}{\kilo\hertz}, respectively. The frequency intervals for the analysis are indicated in fig.~\ref{fig:TEXTOR-117870}b for a medium poloidal distance of $\Delta z=$\SI{11}{\milli\meter}. The poloidal distance, probed by the six different antenna configurations, ranges from \SIrange{4}{22}{\milli\meter}. For the calculation of the mean $v_\perp$, only the five combinations with $\Delta z\ge$\SI{6}{\milli\meter} are used to avoid disturbances by broad-band turbulence which perturb the measurements for smaller probing distances. For the low-frequency turbulence, $v_\perp=$\SI{-1.2}{\kilo\meter\per\second} is deduced. The QC-mode rotation yields $v_{QC}=$\SI{-2.3}{\kilo\meter\per\second}. The error bar for both measurements amounts to \SI{\pm 0.3}{\kilo\meter\per\second}. As for W7-X, the QC-mode rotation is $\approx 2$ times faster than the $E\times B$-rotation, described by the low-frequency turbulence, and the QC-mode is rotating in electron diamagnetic drift direction. As outlined in Eqn.~\ref{equ:mode-number}, the calculated poloidal structure size with $L_{QC}=$\SI{27}{\milli\meter} is within the range expected for TEMs.\\

\begin{figure}
    \centering
    \includegraphics[scale=0.8]{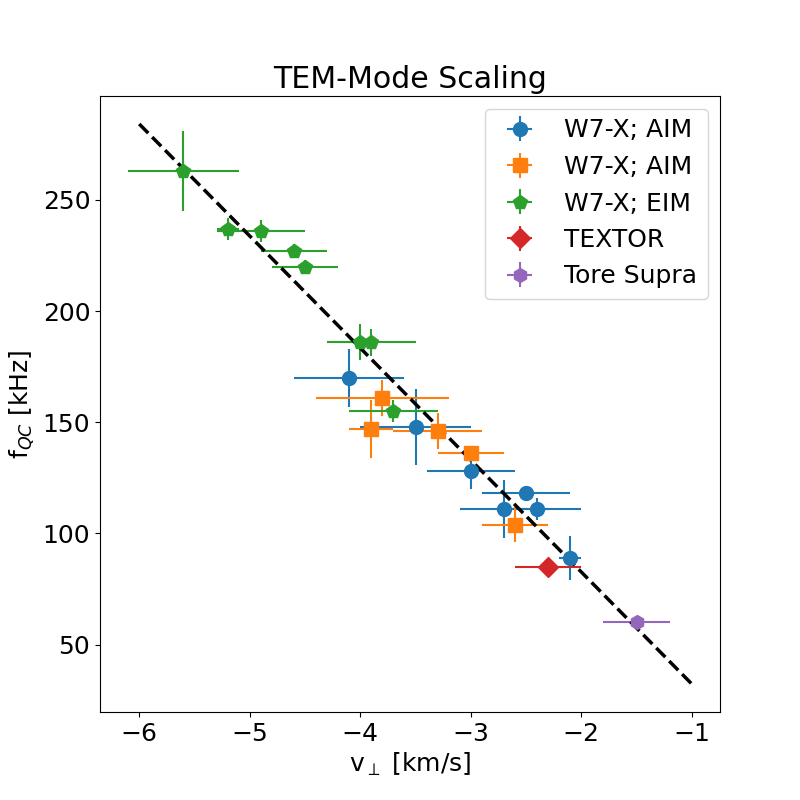}
    \caption{Scaling of the QC-mode frequency as function of the QC-mode rotation for different magnetic configurations. The dashed line indicates the slope of the linear relation from the W7-X data sets and is a measure for the poloidal size of the structure. Furthermore, two data points for the tokamaks TEXTOR and Tore Supra are shown, thus they obey the same scaling.}
    \label{fig:QC-mode_Scaling}
\end{figure}

\noindent So far, the data suggests a linear relation between the QC-mode velocity and the measured centre frequency $f_{QC}$. In fig.~\ref{fig:QC-mode_Scaling}, the scaling is shown for three W7-X programs. It is observed that the QC-mode frequency increases with the QC-mode velocity and the slope of this linear function is a measure for the poloidal structure length of the QC-mode. In the frequency scaling from W7-X the data for TEXTOR extrapolate the scaling to lower $v_{QC}$ and $f_{QC}$ and are in good agreement with the scaling. Also, the data point from Tore Supra, which extrapolates the frequency scaling to even lower frequencies, fits well. Moreover, for this point nonlinear gyrokinetic simulations as well as synthetic diagnostic have been applied, both showing the $\nabla T_e$-TEM nature of the QC-mode in Tore Supra. This scaling seems to be robust and combines QC-mode investigations in low-collisionality plasmas in stellarators and tokamaks. 

\section{Summary and Conclusions}\label{Sum}
\noindent This paper reports on the first observation and interpretation of quasi-coherent modes in the stellarator W7-X. The QC-modes are observed in low-collisionality ECRH-heated plasmas in different magnetic configurations. The QC-mode frequency is found to depend on the magnetic configuration. The mode is observed in the plasma core and its frequency and velocity depends on the injected power. It is found that the QC-mode has an electrostatic nature, because they are not observed in the Mirnov coil diagnostic. The mode observed in the Mirnov  coil diagnostic has an Alfv\'enic nature and shows no dependence on the power. The QC-modes dominate the coherence spectra in the plasma core, as they are obtained from the Poloidal Correlation Reflectometer. The large frequency of the QC-mode allows discriminating mode rotation and $E\times B$-rotation and therefore for both components the poloidal velocity can be  estimated. The QC-mode rotates in electron diamagnetic drift direction and a factor 2 faster than the $E\times B$-rotation, depending on the magnetic configuration. The poloidal structure of the Mode is analysed and sizes of \SIrange{14}{22}{\milli\meter} are obtained, which corresponds to a wave number of $\approx$\SI{3}{\per\centi\meter}. Also, the $k_\perp\rho_s$-value, relevant for the characterization of the transport mechanism, yields values $\ge$\SI{1}{}. This is an indication for trapped-electron-mode turbulence. Linear gyrokinetic simulations performed for the parameters of a W7-X plasma support this hypothesis.\\
\noindent From the measured QC-mode frequency and velocity, a linear scaling of the mode frequency as function of its velocity is suggested. The slope of the scaling is a measure for the poloidal size of the mode. Measurements performed at the tokamaks Tore Supra and TEXTOR that show QC-modes in ohmically heated plasmas are analysed. The calculated mode velocity and the measured frequency of the QC-mode are found to be in agreement with the scaling obtained for W7-X, suggesting that the scaling could be universal. It also adds  evidence on the interpretation of TEMs at W7-X, because at Tore Supra the QC-mode is verified as a TEM. From the frequency scaling, it is possible to identify QC-mode observation at different fusion devices with respect to the underlying transport mechanism. This could be of help in characterizing QC-mode observation in fusion devices in general.\\A remaining task in future investigations is the understanding of the frequency width of the QC-mode. One assumption for the broad structure of the QC-mode is a superposition of many modes, generating the observed QC-mode. They may be generated from trapped particles at different depth of the magnetic well, which could have a discontinuous spectrum. Also, the population of trapped particles in the well could influence the QC-mode width. Another reason for broad frequency spectrum of the QC-mode may be the interaction of the QC-mode with the broadband turbulence, which is supposed to be driven by ion temperature gradients.

\section*{Acknowledgements}
\noindent This work has been carried out within the framework of the EUROfusion Consortium, funded by the European Union via the Euratom Research and Training Programme (Grant Agreement No 101052200 –EUROfusion). Views and opinions expressed are however those of the author(s) only and do not necessarily reflect those of the European Union or the European Commission. Neither the European Union nor the European Commission can be held responsible for them.

\newpage
\section*{References}
\bibliography{literature}
\bibliographystyle{unsrt}

\newpage

\newpage

\newpage

\newpage

\newpage

\newpage

\newpage

\newpage

\newpage

\newpage

\newpage

\newpage

\newpage

\newpage

\newpage

\newpage

\newpage

\newpage

\newpage

\newpage

\newpage

\newpage

\newpage

\newpage

\end{document}